\documentclass[allclo,dvips]{FBSart}
\usepackage{amsfonts}
\usepackage{amssymb}
\usepackage{graphicx} 

\newcommand{\sfrac}[2]{\mbox{\footnotesize $\displaystyle \frac{#1}{#2}$}}
 
\newcommand{\lsim}{\mathrel{\rlap{\lower4pt\hbox{\hskip0pt$\sim$}} 
\raise1pt\hbox{$<$}}}           
\newcommand{\gsim}{\mathrel{\rlap{\lower4pt\hbox{\hskip0pt$\sim$}} 
\raise1pt\hbox{$>$}}}           

\usepackage{color}
\definecolor{purple}{rgb}{0.5,0,0.5}
\definecolor{blue}{rgb}{0.0,0,0.9}

\title{Current quark mass dependence of nucleon magnetic moments and radii}

\author{I.\,C.~Clo\"et,\instnr{1}
G.~Eichmann,\instnr{1,2} 
V.\,V. Flambaum,\instnr{3,4} 
C.\,D.\ Roberts,\instnr{1} 
M.\,S.~Bhagwat\instnr{1}\footnote[4]{Current address: Department of Radiation Oncology, DF/BWH Cancer Center, Harvard Medical School, Boston, MA 02115} and A.\ H\"oll.\instnr{5}\footnote[7]{Current address: BMWi, Villemombler Str.\ 76, D-53123 Bonn, Germany}} 

\instlist{
Physics Division, Argonne National Laboratory, Argonne, IL
60439, USA\and
Institut f\"ur Physik, Karl-Franzens-Universit\"at Graz, A-8010 Graz, Austria\and
Argonne Fellow, Physics Division, Argonne National Laboratory, Argonne, IL
60439, USA\and
School of Physics, University of New South Wales, Sydney 2052, Australia \and
Institut f\"ur Physik, Universit\"at Rostock, D-18051 Rostock, 
Germany}

\runningauthor{V.\,V. Flambaum, et al.}
\runningtitle{Variation of nucleon magnetic moments and radii}
\begin{document}

\maketitle 
\begin{abstract}
A calculation of the current-quark-mass-dependence of nucleon static electromagnetic properties is necessary in order to use observational data as a means to place constraints on the variation of Nature's fundamental parameters.
A Poincar\'e covariant Faddeev equation, which describes baryons as composites of confined-quarks and -nonpointlike-diquarks, is used to calculate this dependence.
The results indicate that, like observables dependent on the nucleons' magnetic moments, quantities sensitive to their magnetic and charge radii, such as the energy levels and transition frequencies in Hydrogen and Deuterium, might also provide a tool with which to place limits on the allowed variation in Nature's constants.  
%


\end{abstract}

\section{Introduction}
It is a feature anticipated of models for the unification of all interactions that the so-called fundamental ``constants'' actually exhibit spatial and temporal variation.  In consequence there is an expanding search for this variation via astronomical, geochemical and laboratory measurements \cite{Uzan:2002vq,Flambaum:2007my}.  An interpretation of these measurements can materially benefit from calculations of the current-quark-mass-dependence of observables characterising hadronic and nuclear systems.  

One example is found in the behaviour of hadron masses \cite{Flambaum:2005kc,Holl:2005st}.  A variation in light-meson masses modifies the internucleon potential, and a variation in the nucleon's mass affects the kinetic energy term in the nuclear Hamiltonian.  Such changes modify nuclear binding energies and can thereby have a substantial impact on Big Bang Nucleosynthesis \cite{Flambaum:2007mj}.  These results enable the use of observational data to place constraints on the variation of Nature's constants; e.g., Ref.\,\cite{Dmitriev:2003qq}.  

Other effects of a variation in hadron masses are an alteration in the location of energy levels and in the positions of compound resonances in heavy nuclei.  For example, Refs.\,\cite{Flambaum:2006ak,FW} explore the sensitivity to changes in the light-quark masses, $m$, of the \emph{nuclear clock} transition between the ground- and first-excited states in $^{229}$Th and the position of the $0.1\,$eV compound resonance in $^{150}$Sm.  It is noteworthy that the shift of the Sm resonance, as determined from the Oklo natural nuclear reactor, currently provides the best terrestrial limit on the temporal variation of Nature's fundamental parameters; namely, $|\dot{X}_q/X_q|<1.6 \times 10^{-18}\,$y$^{-1}$, $X_q:= m/\Lambda_{\rm QCD}$ \cite{Flambaum:2002de,Dmitriev:2002kv,Flambaum:2002wq}.


Hadron magnetic moments can also depend upon current-quark mass.  Calculations of this dependence are necessary for the interpretation of measurements of quasar absorption spectra and superprecise atomic clocks in terms of the variation of Nature's fundamental parameters \cite{Flambaum:2004tm,Tzanavaris:2004bx,Flambaum:2006ip}.  It is notable that in rudimentary constituent-quark models for hadron bound-states composed of degenerate light-quarks, in which the constituents are weakly bound and hence the bound-state's mass is accurately approximated as the sum of constituent-quark masses, the bound-states' magnetic moments are independent of quark mass.  Any quark-mass variation in hadron magnetic moments is therefore a gauge of hadron structure.

Herein we report a calculation of the mass-dependence of neutron and proton magnetic moments based on a Poincar\'e covariant Faddeev equation model for the nucleon.  Since the magnetic and electric form factors are obtained simultaneously, we also describe the variation of the nucleons' magnetic and charge radii.  The background material for our calculation is provided in Sects.\,\ref{nucleonmodel} and \ref{sec:MMR}, and in appendices.  Our results are presented and discussed in Sect.\,\ref{sec:RD}.  Section~\ref{epilogue} is a brief summation.

\section{Nucleon Model}
\label{nucleonmodel}
In quantum field theory a nucleon appears as a pole in a six-point quark Green function.  The pole's residue is proportional to the nucleon's Faddeev amplitude, which is obtained from a Poincar\'e covariant Faddeev equation that adds-up all possible quantum field theoretical exchanges and interactions that can take place between three dressed-quarks.

\begin{figure}[t]
\centerline{%
\includegraphics[clip,width=0.75\textwidth]{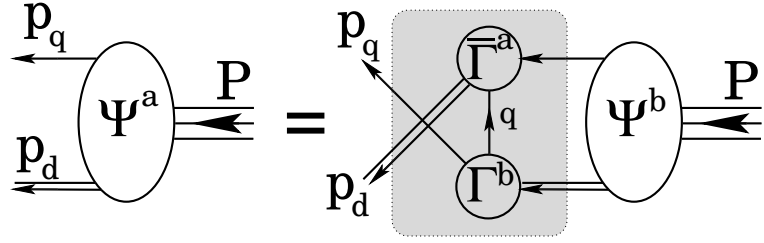}}
\caption{\label{faddeevfigure} Poincar\'e covariant Faddeev equation, Eq.\,(\protect\ref{FEone}), employed herein to calculate nucleon properties.  $\Psi$ in Eq.\,(\protect\ref{PsiNucleon}) is the Faddeev amplitude for a nucleon of total momentum $P= p_q + p_d$.  It expresses the relative momentum correlation between the dressed-quark and -diquarks within the nucleon.  The shaded region demarcates the kernel of the Faddeev equation, Sect.\,\protect\ref{completing}, in which: the \emph{single line} denotes the dressed-quark propagator, Sect.\,\protect\ref{subsubsec:S}; $\Gamma$ is the diquark Bethe-Salpeter-like amplitude, Sect.\,\protect\ref{qqBSA}; and the \emph{double line} is the diquark propagator, Sect.\,\protect\ref{qqprop}.}
\end{figure}

A tractable truncation of the Faddeev equation is based \cite{Cahill:1988dx} on the observation that an interaction which describes mesons also generates diquark correlations in the colour-$\bar 3$ channel \cite{Cahill:1987qr}.  The dominant correlations for ground state octet and decuplet baryons are scalar ($0^+$) and axial-vector ($1^+$) diquarks because, for example: the associated mass-scales are smaller than the baryons' masses \cite{Burden:1996nh,Maris:2002yu}, namely (in GeV)
\begin{equation}
\label{diquarkmass}
m_{[ud]_{0^+}} = 0.7 - 0.8
 \,,\; 
m_{(uu)_{1^+}}=m_{(ud)_{1^+}}=m_{(dd)_{1^+}}=0.9 - 1.0\,;
\end{equation}
and the electromagnetic size of these correlations \cite{Maris:2004bp}
\begin{equation}
r_{[ud]_{0^+}} \approx 0.7\,{\rm fm}\,,\; r_{(ud)_{1^+}} \sim 0.8\,{\rm fm}\,,
\end{equation}
is less than that of the proton.  (The last result is an estimate based on the $\rho$-meson/$\pi$-meson radius-ratio \cite{Maris:2000sk,Bhagwat:2006pu}.)

The kernel of the Faddeev equation is completed by specifying that the quarks are dressed, with two of the three dressed-quarks correlated always as a colour-$\bar 3$ diquark.  As illustrated in Fig\,\ref{faddeevfigure}, binding is then effected by the iterated exchange of roles between the bystander and diquark-participant quarks.  

The Faddeev equation is detailed in \ref{app:FE}~\textit{Faddeev Equation}.  With all its elements specified, as described therein, the equation can be solved to obtain the nucleon's mass and amplitude.  Owing to Eq.\,(\ref{DQPropConstr}), in this calculation the masses of the scalar and axial-vector diquarks are the only variable parameters.  The axial-vector mass is chosen so as to obtain a desired mass for the $\Delta$,\footnote{This is natural because the spin- and isospin-$3/2$ $\Delta$ contains only an axial-vector diquark.  The relevant Faddeev equation is not different in principle to that for the nucleon.  It is described e.g.\ in Ref.\,\protect\cite{Flambaum:2005kc}.} and the scalar mass is subsequently set by requiring a particular nucleon mass.  

We have written here of desired rather than experimental mass values because it is known that the masses of the nucleon and $\Delta$ are materially reduced by pseudoscalar meson cloud effects.  This is discussed in detail in Refs.\,\cite{Hecht:2002ej,Young:2002cj}.  Hence, a baryon represented by the Faddeev equation described above must possess a mass that is inflated with respect to experiment so as to allow for an additional attractive contribution from the pseudoscalar cloud.  As in previous work \cite{Flambaum:2005kc,Alkofer:2004yf} and reported in Table~\ref{tableNmass}, we require $M_N=1.18\,$GeV and $m_\Delta=1.33\,$GeV.  The results and conclusions of our study are essentially unchanged should even larger masses and a smaller splitting $M_\Delta-M_N$ be more realistic, a possibility suggested by Ref.\,\cite{AWTErice}.

\begin{table}[t]
\begin{center}
\caption{\label{tableNmass} Mass-scale parameters (in GeV) for the scalar and axial-vector diquark correlations, fixed by fitting nucleon and $\Delta$ masses offset to allow for ``pion cloud'' contributions \protect\cite{Hecht:2002ej}.  We also list $\omega_{J^{P}}= \sfrac{1}{\surd 2}m_{J^{P}}$, which is the width-parameter in the $(qq)_{J^P}$ Bethe-Salpeter amplitude, Eqs.\,(\protect\ref{Gamma0p}) \& (\protect\ref{Gamma1p}):  its inverse is an indication of the diquark's matter radius.  Row~3 illustrates effects of omitting the axial-vector diquark correlation: the $\Delta$ cannot be formed and $M_N$ is significantly increased.  Evidently, the axial-vector diquark provides significant attraction in the Faddeev equation's kernel.}
\begin{tabular*}{1.0\textwidth}{
c@{\extracolsep{0ptplus1fil}}c@{\extracolsep{0ptplus1fil}}
c@{\extracolsep{0ptplus1fil}} c@{\extracolsep{0ptplus1fil}}c@{\extracolsep{0ptplus1fil}}c@{\extracolsep{0ptplus1fil}}}
\hline
$M_N$ & $M_{\Delta}$~ & $m_{0^{+}}$ & $m_{1^{+}}$~ &
$\omega_{0^{+}} $ & $\omega_{1^{+}}$ \\
\hline
1.18 & 1.33~ & 0.796 & 0.893 & 0.56=1/(0.35\,{\rm fm}) & 0.63=1/(0.31\,{\rm fm}) \\
1.46 &  & 0.796 &  & 0.56=1/(0.35\,{\rm fm}) &  \\
\hline
\end{tabular*}
\end{center}
\end{table}

\section{Magnetic Moments and Charge Radii}
\label{sec:MMR}
\subsection{Background}
The nucleon's electromagnetic current is
\begin{eqnarray}
\label{Jnucleon}
J_\mu(P^\prime,P) & = & ie\,\bar u(P^\prime)\, \Lambda_\mu(q,P) \,u(P)\,, \\
& = &  i e \,\bar u(P^\prime)\,\left( \gamma_\mu F_1(Q^2) +
\frac{1}{2M}\, \sigma_{\mu\nu}\,Q_\nu\,F_2(Q^2)\right) u(P)\,,
\label{JnucleonB}
\end{eqnarray}
where $P$ ($P^\prime$) is the momentum of the incoming (outgoing) nucleon, $Q= P^\prime - P$, and $F_1$ and $F_2$ are, respectively, the Dirac and Pauli form factors.  They are the primary calculated quantities, from which one obtains the nucleon's electric and magnetic (Sachs) form factors 
\begin{equation}
\label{GEpeq}
G_E(Q^2)  =  F_1(Q^2) - \frac{Q^2}{4 M^2} F_2(Q^2)\,,\;  
G_M(Q^2)  =  F_1(Q^2) + F_2(Q^2)\,.
\end{equation}
The nucleons' magnetic moments are defined through
\begin{equation}
\label{momdef}
\mu_n = \kappa_n = G_M^n(0)\,, \; \mu_p = 1 + \kappa_p = G_M^p(0)\,,
\end{equation}
where $\kappa_N$, $N=n,p$, are referred to as the anomalous magnetic moments.  Of course, the nucleon's electric charges, $G_E^N(0)$, are conserved and cannot depend on the current-quark mass.  That is not true of their electric and magnetic radii:
\begin{eqnarray}
\label{chargeradii}
r_p^2 &:=&  -6 \left. \frac{d}{ds} G_E^p(s) \right|_{s=0} ,\; 
r_n^2 := -6 \left. \frac{d}{ds} \, G_E^n(s) \right|_{s=0} ,
\\
(r_N^\mu)^2 &:=& -6 \left. \frac{d}{ds} \ln G_M^N(s) \right|_{s=0}.
\end{eqnarray}

\begin{figure}[t]
\begin{minipage}[t]{\textwidth}
\begin{minipage}[t]{0.45\textwidth}
\leftline{\includegraphics[width=0.90\textwidth]{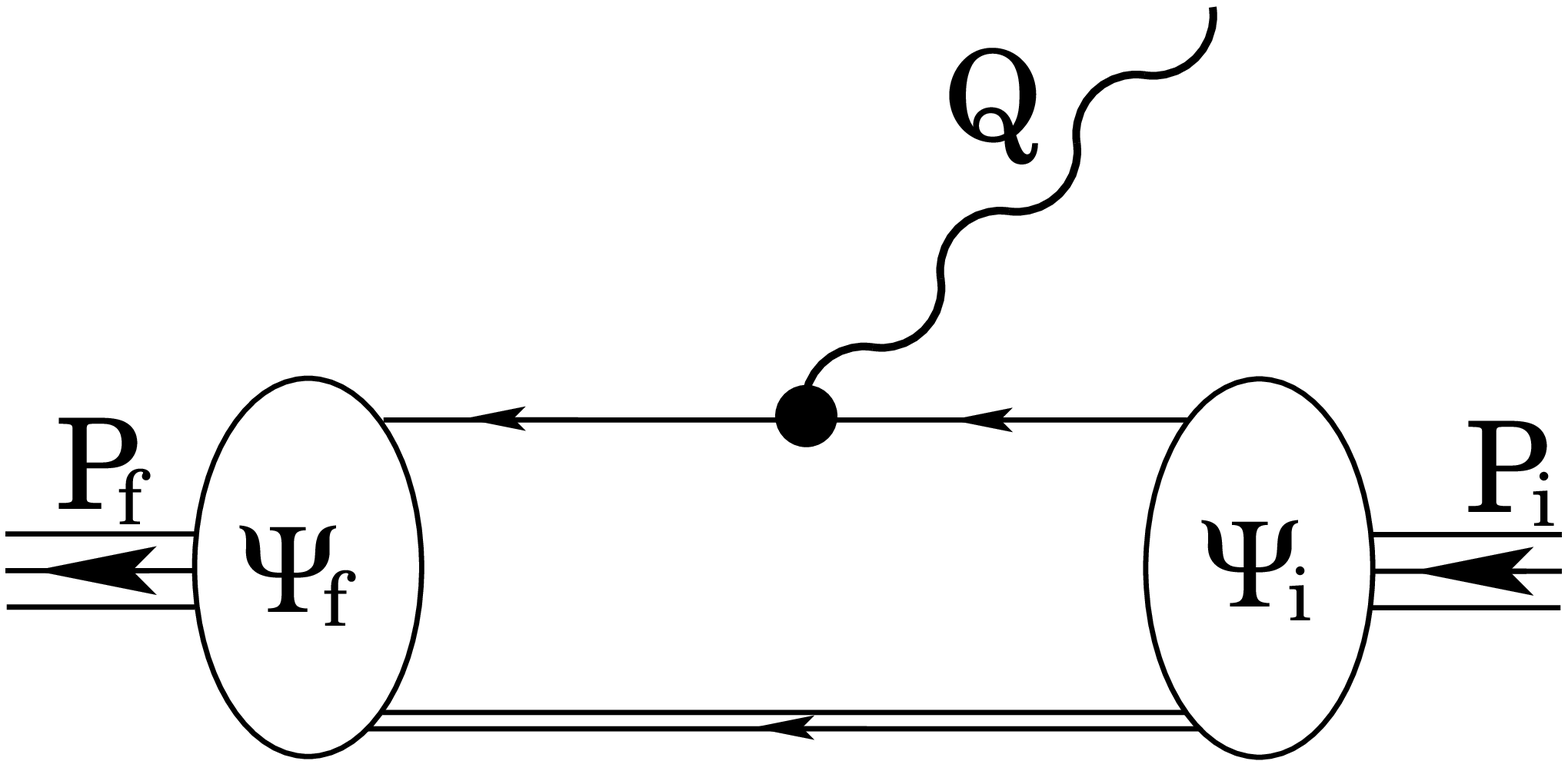}}
\end{minipage}
\begin{minipage}[t]{0.45\textwidth}
\rightline{\includegraphics[width=0.90\textwidth]{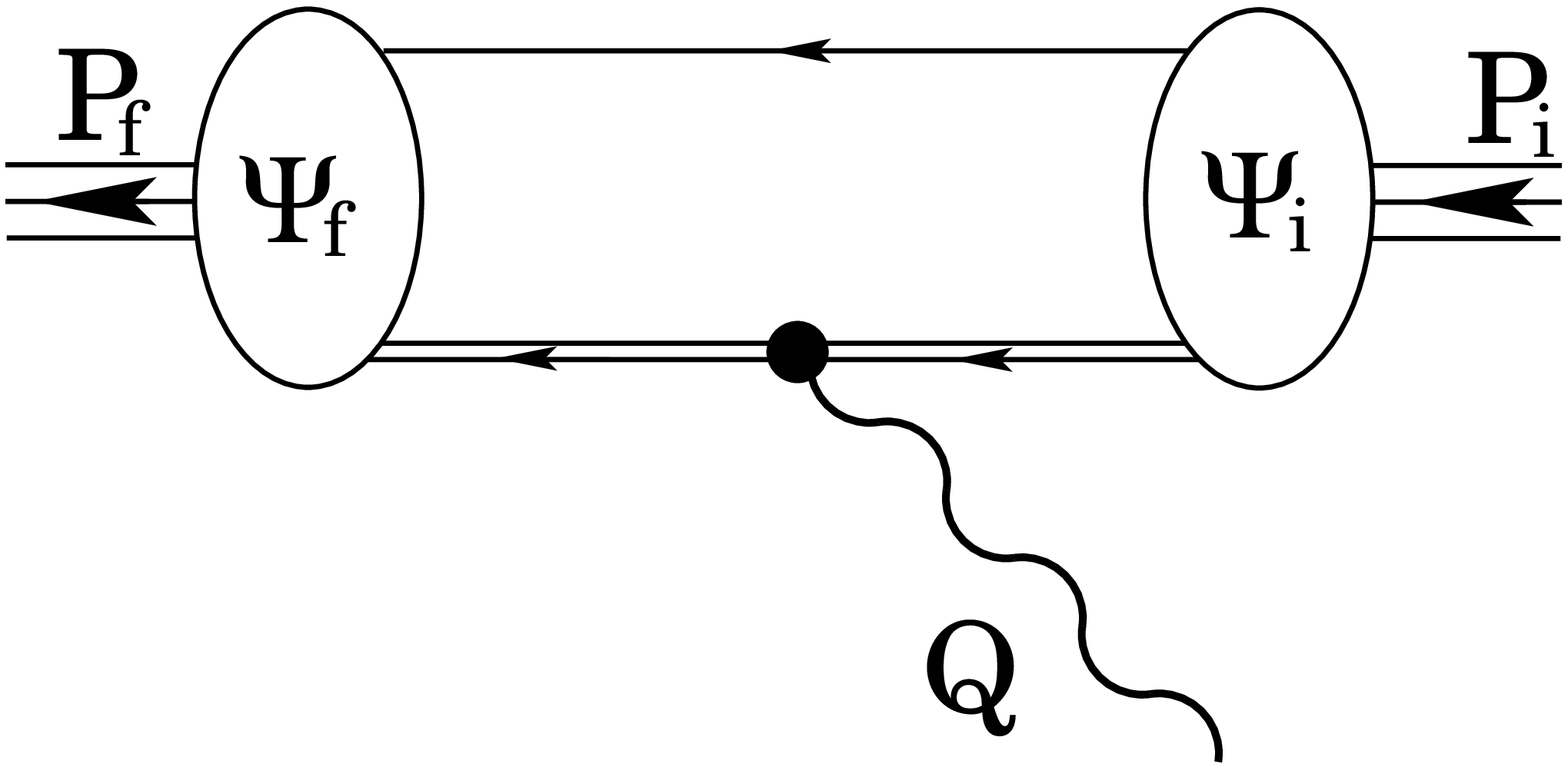}}
\end{minipage}\vspace*{3ex}

\begin{minipage}[t]{0.45\textwidth}
\leftline{\includegraphics[width=0.90\textwidth]{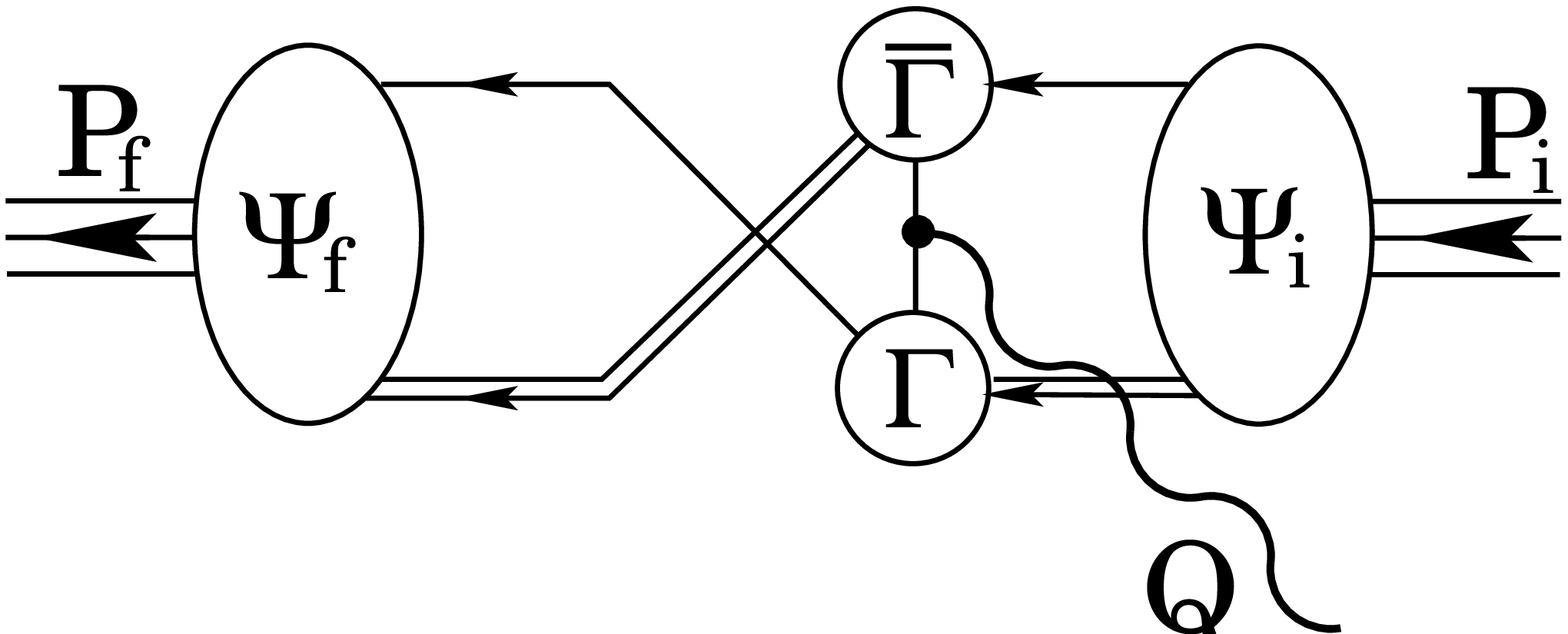}}
\end{minipage}
\begin{minipage}[t]{0.45\textwidth}
\rightline{\includegraphics[width=0.90\textwidth]{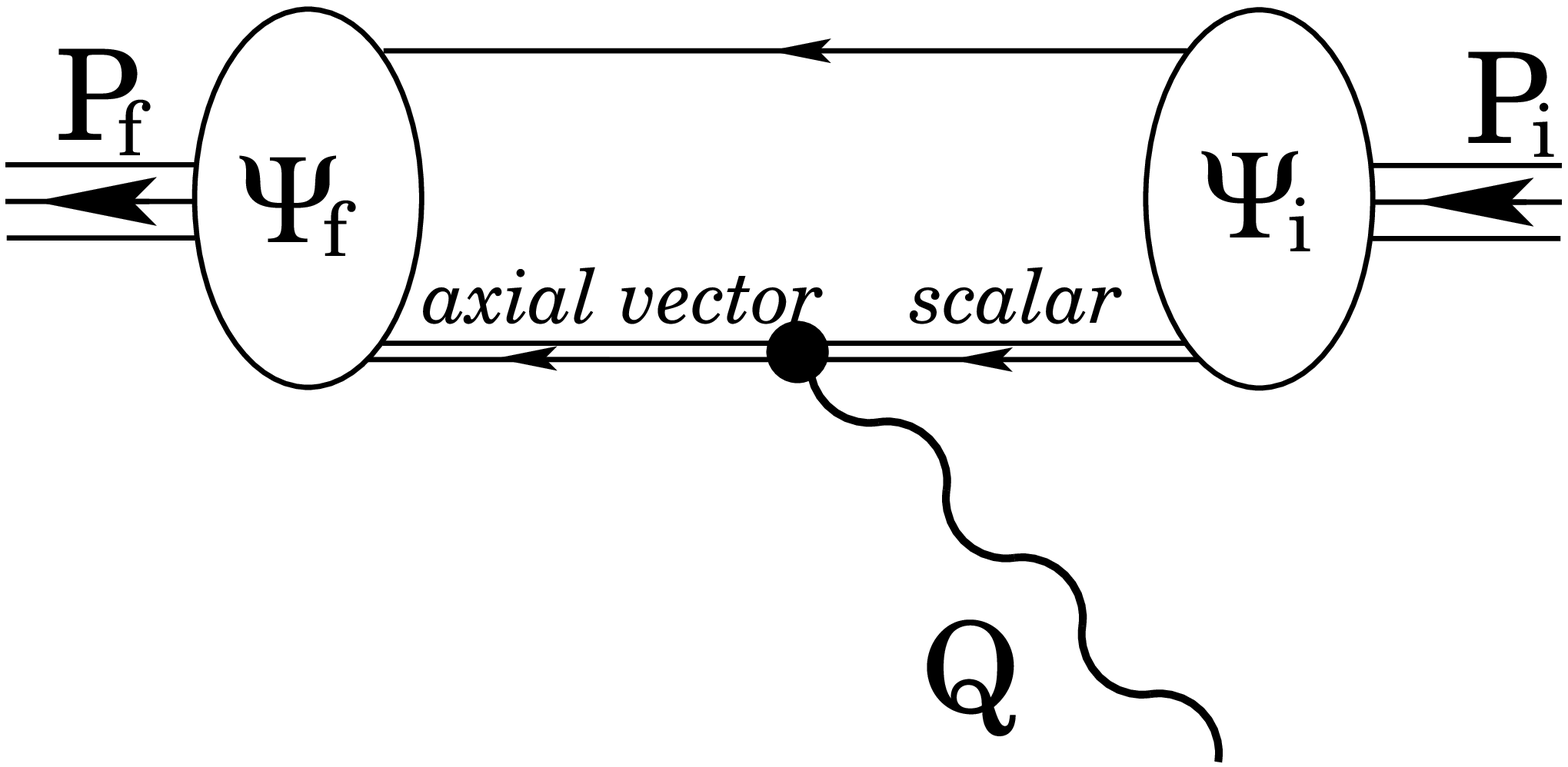}}
\end{minipage}\vspace*{3ex}

\begin{minipage}[t]{0.45\textwidth}
\leftline{\includegraphics[width=0.90\textwidth]{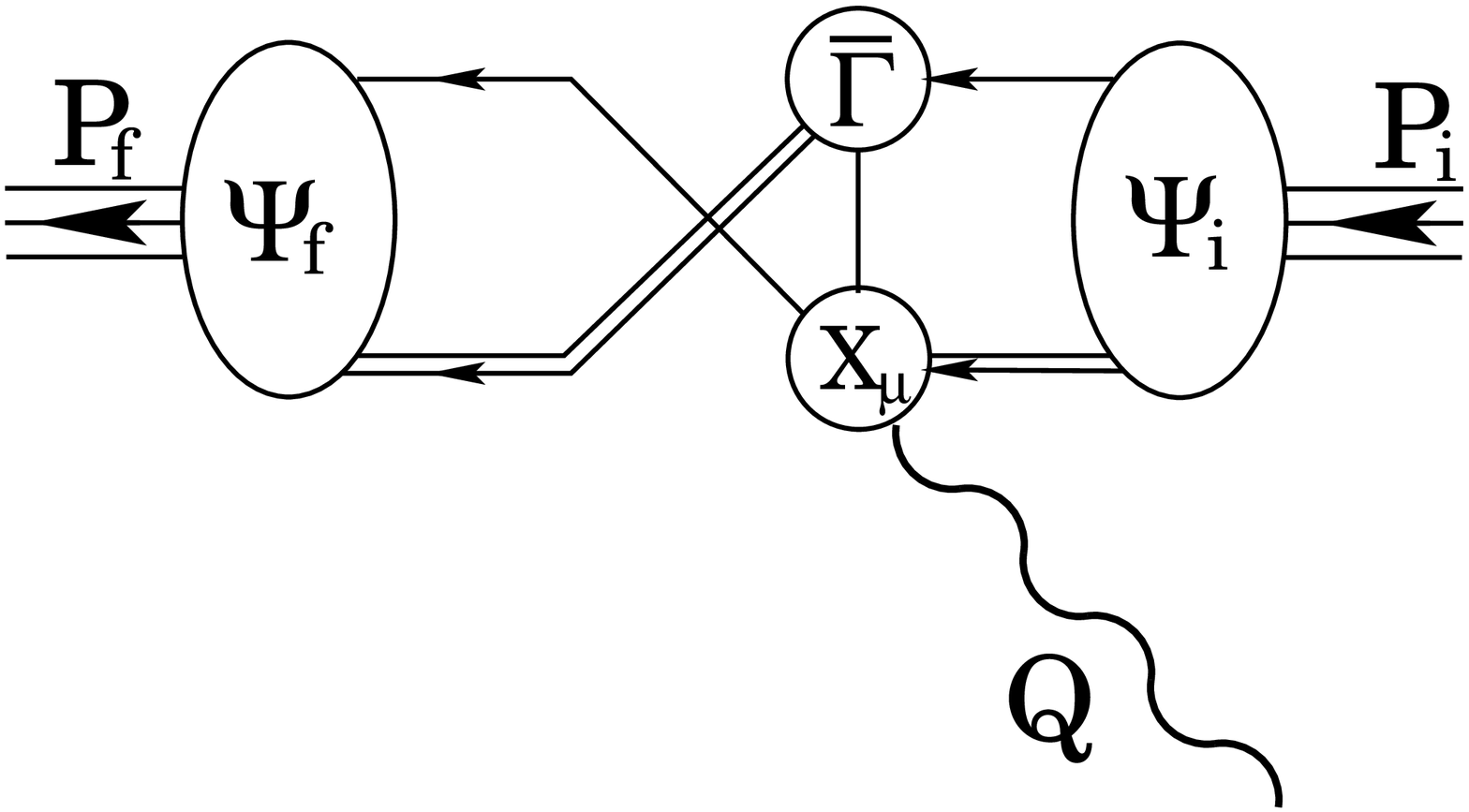}}
\end{minipage}
\begin{minipage}[t]{0.45\textwidth}
\rightline{\includegraphics[width=0.90\textwidth]{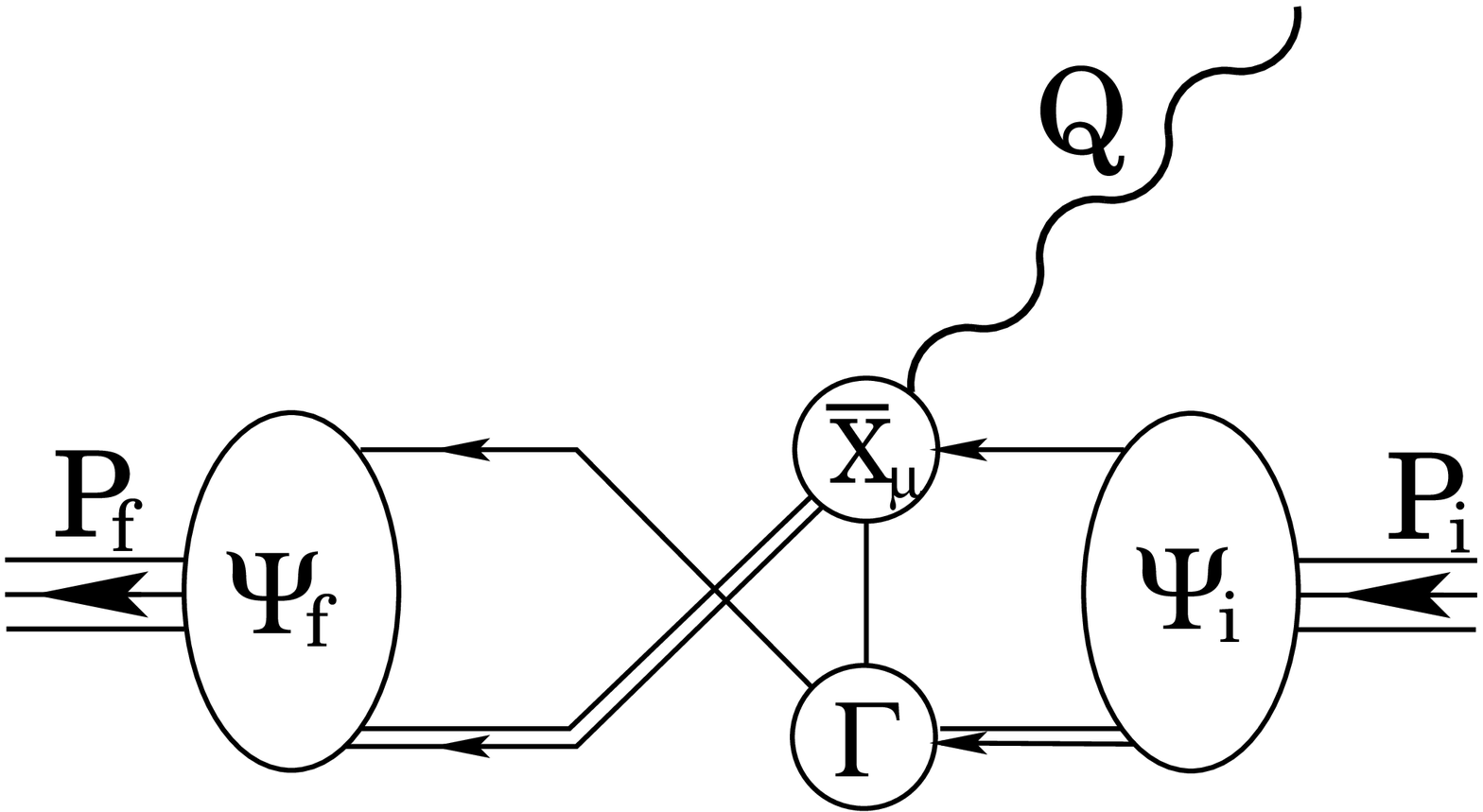}}
\end{minipage}
\end{minipage}
\caption{\label{vertex} Vertex which ensures a conserved current for on-shell nucleons described by the Faddeev amplitudes, $\Psi_{i,f}$, described in Sect.\,\protect\ref{nucleonmodel} and \protect\ref{app:FE}~\emph{Faddeev Equation}.  The single line represents $S(p)$, the dressed-quark propagator, Sec.\,\protect\ref{subsubsec:S}, and the double line, the diquark propagator, Sec.\,\protect\ref{qqprop}; $\Gamma$ is the diquark Bethe-Salpeter amplitude, Sec.\,\protect\ref{qqBSA}; and the remaining vertices are described in \ref{NPVertex} the top-left image is Diagram~1; the top-right, Diagram~2; and so on, with the bottom-right image, Diagram~6.}
\end{figure}

In order to calculate the magnetic moments and charge radii, and their dependence on current-quark mass, one must know the manner in which the nucleon described in Sect.\,\ref{nucleonmodel} couples to a photon.  That is derived in Ref.\,\cite{Oettel:1999gc}, illustrated in Fig.\,\ref{vertex} and detailed in \ref{NPVertex}~\emph{Nucleon-Photon Vertex}.  Naturally, as apparent in that Appendix, the current depends on the electromagnetic properties of the diquark correlations.

\subsection{Pseudoscalar meson loops}
The framework we have described hitherto provides what might be called the quark-core contribution to the nucleons' electromagnetic form factors.  As with the mass, the nucleons' magnetic moments, and charge and magnetic radii receive material contributions from the pseudoscalar meson cloud \cite{radiiCh,young}.  There are two types of contribution: regularisation-scheme-dependent terms, which are analytic functions of $m$ in the neighbourhood of vanishing current-quark mass, $m = 0$; and nonanalytic scheme-independent terms.  

For the magnetic moments and radii the leading-order scheme-independent contributions are \cite{kubis}
\begin{eqnarray}
\label{mpnpion}
(\mu_{n/p})_{NA}^{\rm 1-loop} &\stackrel{m_\pi \simeq 0}{=}& \pm \, \frac{g_A^2\, M_N}{4\pi^2 f_\pi^2}\, m_\pi\,,\\
\label{rpnpion}
\langle r_{n/p}^2\rangle^{1-loop}_{NA} &\stackrel{m_\pi \simeq 0}{=}& \pm\,\frac{1+5 g_A^2}{32 \pi^2 f_\pi^2} \,\ln (\frac{m_\pi^2}{M_N^2}) \,, \\
\label{rpnmpion}
\langle (r_{N}^\mu)^2\rangle^{1-loop}_{NA} &\stackrel{m_\pi \simeq 0}{=}& -\,\frac{1+5 g_A^2}{32 \pi^2 f_\pi^2} \,\ln (\frac{m_\pi^2}{M_N^2})+ \frac{g_A^2\, M_N}{16 \pi f_\pi^2 \mu_v} \frac{1}{m_\pi} \,,
\end{eqnarray}
where, experimentally, $g_A=1.26$, $f_\pi=0.0924$\,GeV\,$=1/(2.13 \,{\rm fm})$ and $\mu_V=\mu_p-\mu_n=4.7$.  These terms reduce the magnitude of both neutron and proton magnetic moments, and increase the magnitudes of the radii.

While these scheme-independent terms are important, at physical values of the pseudoscalar meson masses they do not usually provide the dominant contribution to observables.  That is provided by the regularisation-parameter-dependent terms, as is apparent for baryon masses in Ref.\,\cite{Hecht:2002ej} and for the pion charge radius in Ref.\,\cite{Alkofer:1993gu}.  This is particularly significant for the magnetic moments, in which connection the regularisation-scheme-dependent terms provide a nonzero contribution in the chiral limit and have the net effect of increasing $|\mu_{N}|$.  

This last fact was overlooked in Ref.\,\cite{Alkofer:2004yf} so that Eq.\,(82) therein is a poor model for the net contribution to $\mu_{n,p}$ from pseudoscalar meson loops.  A minimal improvement is
\begin{equation}
\label{ImproveAH}
(\mu_{n/p})^{{\rm 1-loop}^R} = \left( \mu_{n/p}^{\pi 0} \; \pm \; \frac{g_A^2\, M_N}{4\pi^2 f_\pi^2}\, m_\pi \right) \frac{2}{\pi}\arctan(\frac{\lambda^3}{m_\pi^3})\,,
\end{equation}
where $\mu_{n/p}^{\pi 0}$ are the chiral limit values of the meson loop contributions. Equation~(\ref{ImproveAH}) reproduces Eq.\,(\ref{mpnpion}) but also enables us to express reasonable estimates of the regularisation-parameter-dependent parts of the chiral loops.\footnote{NB.\ The expression in Eq.\,(\protect\ref{ImproveAH}) vanishes when the pion mass is much larger than the regularisation scale.  This is required because very massive states must decouple from low-energy phenomena.}   The parameter $\lambda$ is a mass-scale.  Its presence echoes a physical effect; namely, that finite meson size guarantees an intrinsic regularisation of loop integrals.  Emulating  Ref.\,\cite{Alkofer:2004yf}, we employ $\lambda = 0.3\,$GeV$\,=1/[0.66\, {\rm fm}]$.

A recent estimate from numerical simulations of lattice-regularised QCD \cite{Wang:2007iw} gives the following chiral-loop contributions to the nucleons' magnetic moments at the physical pion mass:
\begin{equation}
\label{latticemupi}
\mu_{n}^{\pi} = -0.40\,, \; \mu_{p}^{\pi} = 0.24\,.
\end{equation}
They are obtained with\footnote{Variations in the regularisation mass-scale have no impact; e.g., using $\lambda=0.5\,$GeV in Eq.\,(\protect\ref{ImproveAH}) alters these results by less-than 2\%.}  
\begin{equation}
\mu_{n}^{\pi 0} = -1.05\,, \; \mu_{p}^{\pi 0} = 0.88\,,
\end{equation}
in Eq.\,(\ref{ImproveAH}).  Equation~(\ref{latticemupi}) suggests that the dressed-quark core we have described above should yield the follow magnetic moments for the nucleons at the physical current-quark mass, Eq.\,(\ref{mcq}):
\begin{equation}
\label{qcore}
\mu_{n}^{q(qq)} = -1.51\,, \; \mu_{p}^{q(qq)} = 2.55\,.
\end{equation}
These values provide the comparison column in Table~\ref{magmoms}.

It was argued in Ref.\,\cite{Alkofer:2004yf} that the dressed-quark core described herein is compatible with augmentation by a sensibly regulated pseudoscalar meson cloud.  Indeed, it was argued strongly that agreement with experimental data should not be achieved in the absence of such contributions.  Given Eq.\,(\ref{ImproveAH}), we reconsider this below.

\subsection{Magnetic properties of axial-vector diquarks}
As explained in \ref{NPVertex}~\emph{Nucleon-Photon Vertex}, the nucleons' electromagnetic current involves three parameters, which characterise the axial-vector diquarks' electromagnetic properties; viz., the diquarks' magnetic moment -- $\mu_{1^+}$, their quadrupole moment -- $\chi_{1^+}$, and the scalar$\,\leftrightarrow\,$axial-vector transition strength --  $\kappa_{\cal T}$.  
Calculations have shown \cite{Alkofer:2004yf} that the quadrupole moment has no material impact.  Its greatest effect is on the neutron's charge radius, Eq.\,(\ref{chargeradii}).  However, that is notoriously sensitive to details of the neutron's Faddeev amplitude, which determine the magnitude and sign of the Dirac form factor's slope, and to meson-related contributions, so that, within the accuracy of the model described herein, the net result for all quantities considered is essentially independent of $\chi_{1^+}\in [0,2]$.  
Variations in the diquarks' magnetic moment and transition strength modify the nucleons' radii by $\lesssim 2$\%, which is insignificant.  Their only real impact is on the nucleons' magnetic moments.

\begin{table}[t]
\begin{center}
\caption{\label{magmoms} Values of the magnetic moments defined in Eq.\,(\protect\ref{momdef}) calculated with the diquark mass values in Table.~\protect\ref{tableNmass} and $\chi_{1^+} = 1$.  Experimental values are \protect\cite{Yao:2006px} $\mu_n = -1.91$, $\mu_p=2.79$.}
\begin{tabular*}{1.0\textwidth}
{c@{\extracolsep{0ptplus1fil}}c@{\extracolsep{0ptplus1fil}}
c@{\extracolsep{0ptplus1fil}}c@{\extracolsep{0ptplus1fil}}}
\hline
        & Eq.\,(\ref{qcore}) -- quark-core 
        & $\mu_{1^+} = 0$ 
        & $\mu_{1^+} = 0.37 =:\mu_{1^+}^{\rm f}$ \rule{0em}{3ex}\\[1ex]
        & target value 
        & $\kappa_{\cal T} =0 $ 
        & $\kappa_{\cal T}=0.12=:\kappa_{\cal T}^{\rm f}$  \\\hline
$\mu_n^{q(qq)}$ & -1.51 & -1.44 & -1.51\\
$\mu_p^{q(qq)}$ & ~2.55 & ~2.39 & ~2.55\\\hline
\end{tabular*}
\end{center}
\end{table}

These observations mean that in order to reconsider the claim reiterated above, it is only necessary to reanalyse the magnetic moment calculations in Ref.\,\cite{Alkofer:2004yf}.
In this case Column~3 of Table~\ref{magmoms} reports that with magnetically inert diquarks the quark-core magnetic moments are underestimated.  Moreover, Column~4 indicates that there are values of the diquarks' magnetic moment and transition strength for which the required quark-core moments, Eq.\,(\ref{qcore}), can accurately be reproduced.  The magnitudes of $\mu_{1^+}$ and $\kappa_{\cal T}$ that effect this are much smaller than those for an on-shell axial-vector, which were the focus of Ref.\,\cite{Alkofer:2004yf}.  This marked suppression is physically reasonable because the diquarks are far from being on-shell within the nucleon, a fact observed explicitly in Ref.\,\cite{Mineo:2002bg}.  Hence, the dominant contributions to the form factors are provided by Diagrams~1 and 3.  

It follows that the dressed-quark core defined herein is compatible with dressing by a sensibly regulated pseudoscalar meson cloud so long as the magnetic interactions of axial-vector diquark correlations within the nucleon are commensurate in magnitude with those of the meson cloud.  (See also Table~\ref{physicalresults}.)
 
\section{Results and Discussion}
\label{sec:RD}
We can now describe the method by which we computed the $m$-dependence of the nucleons' magnetic moments and radii.  

We first solved the Faddeev equation at each of a number of current-quark masses within a domain that includes the physical current-quark mass, $m^{\rm phys}$ given in Eq.\,(\ref{mcq}).  In our Faddeev equation the dependence on current-quark mass appears explicitly in the dressed-quark propagator described in Sect.\,\ref{subsubsec:S} and implicitly in the diquark masses.  At each value of $m$ the dressed-quark propagator is obtained in an obvious manner, Eqs.\,(\ref{ssm}) and (\ref{svm}), while the diquark masses are given by
\begin{equation}
m_{J^P}(m) = m_{J^P}^{\rm phys} + \sigma_{qq} \, (m/m^{\rm phys} - 1)\,,
\end{equation}
with $m_{J^P}^{\rm phys}$ given in Row~2 of Table~\ref{tableNmass} and $\sigma_{qq} \approx 26\,$MeV being the diquarks' $\sigma$-term \cite{Flambaum:2005kc,Holl:2005st}.  These observations prescribe the $m$-dependence of the Faddeev kernel.  Solving the Faddeev equation then provides a range of nucleon Faddeev amplitudes, one for each selected current-quark mass.  We worked with 
\begin{equation}
\label{mdomain}
m = 3\,{\rm MeV} + j \, \delta m\,, \; \delta m = 1\,{\rm MeV}\,, \; j=0,1,\ldots,7.
\end{equation}

The next step was to calculate, at each value of $m$ in Eq.\,(\ref{mdomain}), the nucleons' electric and magnetic form factors on $Q^2\in (0,1]\,$GeV$^2$.  In doing this we also allowed the diquarks' magnetic properties to evolve with current-quark mass. Owing to the fact that $\Gamma^{J^P}_C:= \Gamma^{J^P}C^\dagger$, where $\Gamma^{J^P}$ is a diquark's Bethe-Salpeter amplitude, satisfies approximately the same Bethe-Salpeter equation as a $J^{-P}$ colour-singlet meson \cite{Cahill:1987qr}, we referred to Ref.\,\cite{Bhagwat:2006pu} for guidance in constraining the variation of the diquarks' magnetic parameters.  For the $\rho$-meson we inferred that in the neighbourhood of the physical $u/d$-quark mass
\begin{equation}
\label{marismum}
\mu_\rho(m) = \mu_\rho^{\rm phys} + 0.002 \, (m/m_{\rm phys} - 1)\,, \; \mu_\rho^{\rm phys} = 2.01\,,
\end{equation}
from which one extracts the renormalisation group invariant ratio
\begin{equation}
\frac{\delta \mu_\rho}{\mu_\rho}/\frac{\delta m}{m} = 0.001\,.
\end{equation}
Plainly, the response is weak and (perhaps surprisingly, given Table~\ref{dmudm} herein) the $\rho$-meson magnetic moment increases with increasing current-quark mass.  Based on this analysis, in our calculations we employed
\begin{eqnarray}
\label{mudqm}
\mu_{1^+}(m) & = &\mu_{1^+}^{\rm f}  + \varsigma^\mu \, 0.002 \, (m/m^{\rm phys} - 1) \,, \\
\label{kappadqm}
\kappa_{\cal T}(m) &= & \kappa_{\cal T}^{\rm f}  + \varsigma^\kappa \, 0.002 \, (m/m^{\rm phys} - 1)\,,
\end{eqnarray}
where $\mu_{1^+}^{\rm f}$ and $\kappa_{\cal T}^{\rm f}$ are given in Table~\ref{magmoms}.  Since the response in Eq.\,(\ref{marismum}) is weak and the sign unexpected, we included the parameters $\varsigma^\mu = \pm 1$ and $\varsigma^\kappa = \pm 1$ in Eqs.\,(\ref{mudqm}), (\ref{kappadqm}) so that we could estimate the error associated with employing these formulae.

We judge it worthwhile to dwell a little longer on properties of mesons that are relevant to our discourse.  From Fig.\,7 of Ref.\,\protect\cite{Bhagwat:2006pu} we infer
\begin{equation}
\frac{\delta r_\rho^2}{r_\rho^2}/\frac{\delta m}{m} \approx 
\frac{\delta (r_\rho^\mu)^2}{(r_\rho^\mu)^2}/\frac{\delta m}{m}\approx -0.06\,.
\end{equation}
This dressed-quark core value possesses the same sign and a similar magnitude to the response of the proton's dressed-quark core radii, described subsequently in connection with Table~\protect\ref{dmudm}.  We have also analysed the response of the pion's dressed-quark core.  A quick estimate can be obtained via the Nambu-Jona--Lasinio model, in which \cite{Blin:1987hw}
\begin{equation}
\label{rpi2njl}
r_\pi^2 = \frac{3}{4\pi^2} \frac{1}{f_\pi^2}\,.
\end{equation}
It is straightforward to solve this model's gap equation, and therefrom calculate $f_\pi$ and its dependence on current-quark mass.  We find thereby
\begin{equation}
q_{r_\pi^2}^{\bar q q} := \frac{\delta r_\pi^2}{r_\pi^2}/\frac{\delta m}{m} =-0.05\,.
\end{equation}
The same result is found by evaluating Eq.\,(\ref{rpi2njl}) with the $m$-dependence of $f_\pi$ depicted in Fig.\,3 of Ref.\,\cite{Holl:2004fr}, which is obtained from a rainbow truncation of QCD's gap equation,   Finally, from Fig.\,6 of Ref.\,\cite{Maris:2005tt}, which reports an internally consistent Dyson-Schwinger equation calculation, we infer $q_{r_\pi^2}^{\bar q q}=-0.06$.  A natural scale associated with the response of dressed-quark core radii is now apparent.  

Given Eq.\,(\ref{marismum}) it was not surprising for us to discover that the nucleons' magnetic moments respond slowly to changing $m$.  Therefore very precise results for the form factors are required in order to compute the slope.  In order to achieve the numerical accuracy necessary we made significant modifications to the computer code employed in Ref.\,\cite{Alkofer:2004yf}.  These changes reduced execution times by an order of magnitude, and made practical and reasonable the use of $10^8$ adaptive Monte-Carlo sample points in the evaluation of Diagrams~3, 5 and 6, which are two-loop integrals.  

\begin{table}[t]
\begin{center}
\caption{\label{physicalresults} Quark-core and pseudoscalar loop [Eqs.\,(\protect\ref{ImproveAH}), (\protect\ref{rpnpionR}), (\protect\ref{rpnmpionR})] contributions to the moments and radii, calculated at the physical current-quark mass, Eq.\,(\protect\ref{mcq}).  The radii are listed in fm$^2$. Experimental values are quoted from Ref.\,\protect\cite{Yao:2006px}, where available, and otherwise from Ref.\,\protect\cite{Mergell:1995bf}.}
\begin{tabular*}{1.0\textwidth}
{c@{\extracolsep{0ptplus1fil}}c@{\extracolsep{0ptplus1fil}}c@{\extracolsep{0ptplus1fil}}
c@{\extracolsep{0ptplus1fil}}c@{\extracolsep{0ptplus1fil}}
c@{\extracolsep{0ptplus1fil}}c@{\extracolsep{0ptplus1fil}}}\hline
\rule{0em}{3ex} & $\mu_n$ & $\mu_p$ & $\langle r_n^2 \rangle$ & $\langle r_p^2 \rangle$ & $\langle (r_n^\mu)^2\rangle $ &  $\langle (r_p^\mu)^2\rangle $\\\hline
 experiment & -1.91 & 2.79& $-(0.34)^2$ & $(0.88)^2$ & $(0.89)^2$ & $(0.84)^2$ \\\hline
 $q(qq)$ & -1.51 & 2.55 & $-(0.13)^2$ & $(0.57)^2$ & $(0.51)^2$ & $(0.50)^2$  \\
 $\pi$-loop & -0.40 & 0.24 & $-(0.47)^2$ & $(0.47)^2$ & $(0.61)^2$ & $(0.61)^2$ \\\hline
 total & -1.91 & 2.79 & $-(0.49)^2$& $(0.74)^2$ & $(0.79)^2$ & $(0.79)^2$ \\\hline
\end{tabular*}
\end{center}
\end{table}

Proceeding as described above we arrived at a set of form factors, one at each value of the current-quark mass in Eq.\,(\ref{mdomain}), and thereby the magnetic moments and radii as a function of $m$.  Each quantity, designated generically by ${\cal Q}$, can be expressed in the form 
\begin{equation}
{\cal Q} = {\cal Q}^{q(qq)} + {\cal Q}^\pi
\end{equation}
where ${\cal Q}^{q(qq)}$ is provided by our Faddeev equation results and ${\cal Q}^\pi$ is an estimate of the contribution from pseudoscalar meson loops.\footnote{NB.\ By construction the Faddeev equation is intended to describe a nucleon's dressed-quark core.  It explicitly excludes all diagrams that can appear in the calculation of ${\cal Q}^\pi$.}  
At the physical current-quark mass, Eq.\,(\ref{mcq}), our calculated results for the former are given in Table~\ref{physicalresults}.  To calculate the latter we employed Eq.\,(\ref{ImproveAH}) and \cite{Alkofer:2004yf,ashley}
\begin{eqnarray}
\label{rpnpionR}
\langle r_{n/p}^2\rangle^{1-loop^R} &=& \pm\,\frac{1+5 g_A^2}{32 \pi^2 f_\pi^2} \,\ln (\frac{m_\pi^2}{m_\pi^2+\lambda^2}) \,, \\
\nonumber \langle (r_{N}^\mu)^2\rangle^{1-loop^R} &=& -\,\frac{1+5 g_A^2}{32 \pi^2 f_\pi^2} \,\ln (\frac{m_\pi^2}{m_\pi^2+\lambda^2}) \\
& & + \frac{g_A^2\, M_N}{16 \pi f_\pi^2 \mu_v} \frac{1}{m_\pi} \,
\frac{2}{\pi}\arctan(\frac{\lambda}{m_\pi})\,.
\label{rpnmpionR}
\end{eqnarray}
These values, too, are reported in the Table, from which it is apparent that pseudoscalar meson loops contribute materially to the nucleons' static properties.  As already remarked, the neutron's electric charge radius is particularly sensitive to details of the neutron's Faddeev amplitude and, we see here, to the size of the meson cloud contribution.  In the computation of form factors, that contribution remains significant for $Q^2 \lesssim 2\,$GeV$^2$.  (See, e.g., Ref.\,\cite{sato}.)

\begin{table}[t]
\begin{center}
\caption{\label{dmudm} Calculated variation of the nucleons' magnetic moments and radii. We list $q_{\cal Q}^s:= - \frac{\delta {\cal Q}^s}{{\cal Q}^s}/\frac{\delta m}{m}$, where $s= q(qq)$, $\pi$, t$\,=\,$total$=q(qq)+\pi$, and $f_{\cal Q}=- \frac{\delta {\cal Q}^{q(qq)}}{{\cal Q}^{\rm t}}/\frac{\delta m}{m}$.}
\begin{tabular*}{1.0\textwidth}{
c@{\extracolsep{0ptplus1fil}}c@{\extracolsep{0ptplus1fil}}
c@{\extracolsep{0ptplus1fil}}c@{\extracolsep{0ptplus1fil}}
c@{\extracolsep{0ptplus1fil}}c@{\extracolsep{0ptplus1fil}}
c@{\extracolsep{0ptplus1fil}}c@{\extracolsep{0ptplus1fil}}
}\hline
$q_{\mu_n}^{q(qq)}$ & $q_{\mu_n}^{\pi}$ & $q_{\mu_n}^{\rm t}$ & $f_{\mu_n}$ & 
$q_{\mu_p}^{q(qq)}$ & $q_{\mu_p}^{\pi}$ & $q_{\mu_p}^{\rm t}$ & $f_{\mu_p}$ \\
0.016 & 0.828 & 0.186 & 0.012 & 0.029 & 1.311 & 0.139 & 0.026\\\hline
$q_{r_n^2}^{q(qq)}$ & $q_{r_n^2}^{\pi}$ & $q_{r_n^2}^{\rm t}$ & $f_{r_n^2}$ &
$q_{r_p^2}^{q(qq)}$ & $q_{r_p^2}^{\pi}$ & $q_{r_p^2}^{\rm t}$ & $f_{r_p^2}$ \\
 0.551 &0.477 & 0.483& 0.041 & 0.014 & 0.477 & 0.202 & 0.008\\\hline
$q_{(r_n^\mu)^2}^{q(qq)}$ & $q_{(r_n^\mu)^2}^{\pi}$ & $q_{(r_n^\mu)^2}^{\rm t}$ & $f_{(r_n^\mu)^2}$ &
$q_{(r_p^\mu)^2}^{q(qq)}$ & $q_{(r_p^\mu)^2}^{\pi}$ & $q_{(r_p^\mu)^2}^{\rm t}$& $f_{(r_p^\mu)^2}$ \\
 0.059 & 0.554 & 0.350 & 0.024 & 0.061 & 0.554 & 0.356 & 0.024 \\\hline
\end{tabular*}
\end{center}
\end{table}

The dependence of the magnetic moments and radii on current-quark mass is expressed through
\begin{equation}
\delta {\cal Q} = \left( \frac{d{\cal Q}}{dm}^{q(qq)}  + \frac{d {\cal Q}}{dm}^\pi  \right) \delta m \,.
\end{equation}
We have described in detail the manner in which we calculated the first term.  The second is computed from Eqs.\,(\protect\ref{ImproveAH}), (\protect\ref{rpnpionR}), (\protect\ref{rpnmpionR}) using Eq.\,(\ref{inpion}).  Our results are listed in Table~\ref{dmudm},\footnote{NB.\ The $(\varsigma^\mu,\varsigma^\kappa)$-variation discussed in connection with Eqs.\,(\protect\ref{mudqm}), (\protect\ref{kappadqm}) affects no result by more than 2\%.}
from which it is immediately apparent that the moments and radii all decrease with increasing current-quark mass.  Moreover, for nucleons, which are composed of $u/d$ valence quarks, the response of the contributions from the dressed-quark core to a change in current-quark mass is $8\pm 3$\% of that arising from the pseudoscalar meson cloud.  

In connection with the magnetic moments it is natural to make a comparison with Ref.\,\cite{Flambaum:2004tm}, which reports in Eq.\,(A7): $\mu_0^n=-2.58$, $\mu_0^p=3.48$.  These values can be compared with our results 
\begin{equation}
\mu_n^{q(qq)}+ \mu_n^{\pi 0} = -2.56\,,\; 
\mu_p^{q(qq)}+ \mu_p^{\pi 0} = 3.43\,.
\end{equation}
In addition, Eqs.\,(25), (27) list $q_{\mu_n}^{\rm t}=0.118$, $q_{\mu_p}^{\rm t}=0.087$ and Eqs.\,(A1), (A5), (A6) enable one to calculate $f_{\mu_n}=0.010$, $f_{\mu_p}=0.012$. %
Our estimation of the loop contributions is cruder than that in Ref.\,\cite{Flambaum:2004tm} but, nevertheless, our values are larger by only a factor of $\approx 1.5$.  
Our dressed-quark core contribution to the variation of $\mu_n$ is almost the same, whereas for the proton it is larger by a factor of two.\footnote{\ref{chargesymmetry}~\emph{Charge Symmetry} explains the relative magnitudes of $q_{\mu_n}^{q(qq)}$ and $q_{\mu_p}^{q(qq)}$.}
We emphasise that our results were obtained through an internally consistent calculation performed directly at the physical light-quark current-mass.  In contrast, Ref.\,\cite{Flambaum:2004tm} extrapolated lattice-regularised QCD results obtained at approximately the strange-quark mass.

These comments notwithstanding, our calculations plainly provide an independent confirmation of the magnitude and sign of the effects discussed in Ref.\,\cite{Flambaum:2004tm}.  This agreement supports a view that all the results reported in Table~\ref{dmudm} are reliable estimates.  

\section{Epilogue}
\label{epilogue}
We used a Poincar\'e covariant Faddeev equation model for the dressed-quark core of the nucleon, augmented by a nucleon-photon vertex which automatically fulfills the Ward-Takahashi identity for on-shell nucleons and a rudimentary estimate of the contribution from pseudoscalar meson loops, to obtain insight into the response of the nucleons' static electromagnetic properties to changes in current-quark mass.  

Our key results are discussed in connection with Table~\ref{dmudm} and summarised by the renormalisation group invariant ratios presented here:
\begin{equation}
\label{dmudmF}
\begin{array}{c|cccccc}
{\cal Q}  & \mu_n & \mu_p & r_n^2& r_p^2 & (r_n^\mu)^2 & (r_p^\mu)^2 \\\hline
\rule{0em}{3ex}- \frac{\delta{\cal Q}}{\cal Q}/\frac{\delta m}{m}  & 0.186 & 0.139 & 0.483 & 0.202 & 0.350 & 0.356\\
\end{array}\,.
\end{equation}  
Those for the magnetic moments can assist in constraining the allowed temporal variation of the current-quark mass via, e.g., experiments with atomic clocks \cite{Flambaum:2007my,Flambaum:2006ip} and various astrophysical measurements \cite{Flambaum:2007my,Tzanavaris:2004bx,Drinkwater:1997ab,Murphy:2001nu}.
Our results also suggest that observables dependent on the nucleons' magnetic and charge radii might provide a useful means by which to place limits on the allowed variation in Nature's fundamental parameters.  It is notable, for example, that the calculated energy levels and transition frequencies in Hydrogen and Deuterium, which are some of the most precise theoretical predictions in physics, are quite sensitive to the value of the proton's charge radius \cite{Mohr:2005zz}.  
Moreover, all hyperfine transition frequencies in atomic clocks \cite{Flambaum:2007my,Flambaum:2006ip} and astrophysics \cite{Flambaum:2007my,Tzanavaris:2004bx,Drinkwater:1997ab,Murphy:2001nu} react to a change in nucleonic charge and magnetisation distributions: the former alters electronic wave functions and the latter changes the hyperfine interaction Hamiltonian.

As a byproduct of this study, we arrived at an improved understanding of diquark correlations.  Within the nucleon they are usually far from being on-shell and hence it is a poor approximation to represent, for example, the active magnetic properties of the axial-vector correlations by on-shell values.  

\begin{acknowledge}
We are grateful to B.~El-Bennich, T.~Kl\"ahn and R.\,D.~Young for profitable discussions.
This work was supported by: the Department of Energy, Office of Nuclear Physics, contract no.\ DE-AC02-06CH11357;
the Austrian Science Fund \emph{FWF} under grant no.\ W1203;
and benefited from the facilities of ANL's Computing Resource Center.
\end{acknowledge}

\appendix

\section{Faddeev Equation}
\setcounter{section}{1}
\label{app:FE}
\subsection{General structure}
The nucleon is represented by a Faddeev amplitude
\begin{equation} 
\label{PsiNucleon}
\Psi = \Psi_1 + \Psi_2 + \Psi_3  \,, 
\end{equation} 
where the subscript identifies the bystander quark and, e.g., $\Psi_{1,2}$ are obtained from $\Psi_3$ by a correlated, cyclic permutation of all the quark labels.  We employ the simplest realistic representation of $\Psi$.  The spin- and isospin-$1/2$ nucleon is a sum of scalar and axial-vector diquark correlations:
\begin{equation} 
\label{Psi} \Psi_3(p_i,\alpha_i,\tau_i) = {\cal N}_3^{0^+} + {\cal N}_3^{1^+}, 
\end{equation} 
with $(p_i,\alpha_i,\tau_i)$ the momentum, spin and isospin labels of the 
quarks constituting the bound state, and $P=p_1+p_2+p_3$ the system's total momentum.\footnote{NB.\ Herein we assume isospin symmetry of the strong interaction; i.e., the $u$- and $d$-quarks are indistinguishable but for their electric charge.  This simplifies the form of the Faddeev amplitudes.}  

The scalar diquark piece in Eq.\,(\ref{Psi}) is 
\begin{eqnarray} 
{\cal N}_3^{0^+}(p_i,\alpha_i,\tau_i)&=& [\Gamma^{0^+}(\sfrac{1}{2}p_{[12]};K)]_{\alpha_1 
\alpha_2}^{\tau_1 \tau_2}\, \Delta^{0^+}(K) \,[{\cal S}(\ell;P) u(P)]_{\alpha_3}^{\tau_3}\,,%
\label{calS} 
\end{eqnarray} 
where: the spinor satisfies (\protect\ref{App:EM}~\emph{Euclidean Conventions})
\begin{equation}
(i\gamma\cdot P + M)\, u(P) =0= \bar u(P)\, (i\gamma\cdot P + M)\,,
\end{equation}
with $M$ the mass obtained by solving the Faddeev equation, and it is also a
spinor in isospin space with $\varphi_+= {\rm col}(1,0)$ for the proton and
$\varphi_-= {\rm col}(0,1)$ for the neutron; $K= p_1+p_2=: p_{\{12\}}$,
$p_{[12]}= p_1 - p_2$, $\ell := (-p_{\{12\}} + 2 p_3)/3$; $\Delta^{0^+}$
is a pseudoparticle propagator for the scalar diquark formed from quarks $1$
and $2$, and $\Gamma^{0^+}\!$ is a Bethe-Salpeter-like amplitude describing
their relative momentum correlation; and ${\cal S}$, a $4\times 4$ Dirac
matrix, describes the relative quark-diquark momentum correlation.  (${\cal
S}$, $\Gamma^{0^+}$ and $\Delta^{0^+}$ are discussed in Sect.\,\ref{completing}.)  The colour antisymmetry of $\Psi_3$ is implicit in $\Gamma^{J^P}\!\!$, with the 
Levi-Civita tensor, $\epsilon_{c_1 c_2 c_3}$, expressed via the antisymmetric 
Gell-Mann matrices; viz., defining 
\begin{equation} 
\{H^1=i\lambda^7,H^2=-i\lambda^5,H^3=i\lambda^2\}\,, 
\end{equation} 
then $\epsilon_{c_1 c_2 c_3}= (H^{c_3})_{c_1 c_2}$.  [See 
Eqs.\,(\ref{Gamma0p}), (\ref{Gamma1p}).]

The axial-vector component in Eq.\,(\ref{Psi}) is
\begin{eqnarray} 
{\cal N}^{1^+}(p_i,\alpha_i,\tau_i) & =&  [{\tt t}^i\,\Gamma_\mu^{1^+}(\sfrac{1}{2}p_{[12]};K)]_{\alpha_1 
\alpha_2}^{\tau_1 \tau_2}\,\Delta_{\mu\nu}^{1^+}(K)\, 
[{\cal A}^{i}_\nu(\ell;P) u(P)]_{\alpha_3}^{\tau_3}\,,
\label{calA} 
\end{eqnarray} 
where the symmetric isospin-triplet matrices are 
\begin{equation} 
{\tt t}^+ = \frac{1}{\surd 2}(\tau^0+\tau^3) \,,\; 
{\tt t}^0 = \tau^1\,,\; 
{\tt t}^- = \frac{1}{\surd 2}(\tau^0-\tau^3)\,, 
\end{equation} 
and the other elements in Eq.\,(\ref{calA}) are straightforward generalisations of those in Eq.\,(\ref{calS}). 

The general forms of the matrices ${\cal S}(\ell;P)$, ${\cal A}^i_\nu(\ell;P)$ and ${\cal D}_{\nu\rho}(\ell;P)$, which describe the momentum space correlation between the quark and diquark in the nucleon are described in Refs.\,\cite{Oettel:1998bk,Cloet:2007pi}.  The requirement that ${\cal S}(\ell;P)$ represent a positive energy nucleon; namely, that it be an eigenfunction of $\Lambda_+(P)$, Eq.\,(\ref{Lplus}), entails
\begin{equation}
\label{Sexp} 
{\cal S}(\ell;P) = s_1(\ell;P)\,I_{\rm D} + \left(i\gamma\cdot \hat\ell - \hat\ell \cdot \hat P\, I_{\rm D}\right)\,s_2(\ell;P)\,, 
\end{equation} 
where $(I_{\rm D})_{rs}= \delta_{rs}$, $\hat \ell^2=1$, $\hat P^2= - 1$.  In the nucleon rest frame, $s_{1,2}$ describe, respectively, the upper, lower component of the bound-state nucleon's spinor.  Placing the same constraint on the axial-vector component, one has
\begin{equation}
\label{Aexp}
 {\cal A}^i_\nu(\ell;P) = \sum_{n=1}^6 \, p_n^i(\ell;P)\,\gamma_5\,A^n_{\nu}(\ell;P)\,,\; i=+,0,-\,,
\end{equation}
where ($ \hat \ell^\perp_\nu = \hat \ell_\nu + \hat \ell\cdot\hat P\, \hat P_\nu$, $ \gamma^\perp_\nu = \gamma_\nu + \gamma\cdot\hat P\, \hat P_\nu$)
\begin{equation}
\begin{array}{lll}
A^1_\nu= \gamma\cdot \hat \ell^\perp\, \hat P_\nu \,,\; &
A^2_\nu= -i \hat P_\nu \,,\; &
A^3_\nu= \gamma\cdot\hat \ell^\perp\,\hat \ell^\perp\,,\\
A^4_\nu= i \,\hat \ell_\mu^\perp\,,\; &
A^5_\nu= \gamma^\perp_\nu - A^3_\nu \,,\; &
A^6_\nu= i \gamma^\perp_\nu \gamma\cdot\hat \ell^\perp - A^4_\nu\,.
\end{array}
\end{equation}

One can now write the Faddeev equation satisfied by $\Psi_3$ as
\begin{equation} 
 \left[ \begin{array}{r} 
{\cal S}(k;P)\, u(P)\\ 
{\cal A}^i_\mu(k;P)\, u(P) 
\end{array}\right]\\ 
 = -\,4\,\int\frac{d^4\ell}{(2\pi)^4}\,{\cal M}(k,\ell;P) 
\left[ 
\begin{array}{r} 
{\cal S}(\ell;P)\, u(P)\\ 
{\cal A}^j_\nu(\ell;P)\, u(P) 
\end{array}\right] .
\label{FEone} 
\end{equation} 
The kernel in Eq.~(\ref{FEone}) is 
\begin{equation} 
\label{calM} {\cal M}(k,\ell;P) = \left[\begin{array}{cc} 
{\cal M}_{00} & ({\cal M}_{01})^j_\nu \\ 
({\cal M}_{10})^i_\mu & ({\cal M}_{11})^{ij}_{\mu\nu}\rule{0mm}{3ex} 
\end{array} 
\right] 
\end{equation} 
with 
\begin{equation} 
 {\cal M}_{00} = \Gamma^{0^+}\!(k_q-\ell_{qq}/2;\ell_{qq})\, 
S^{\rm T}(\ell_{qq}-k_q) \,\bar\Gamma^{0^+}\!(\ell_q-k_{qq}/2;-k_{qq})\, 
S(\ell_q)\,\Delta^{0^+}(\ell_{qq}) \,, 
\end{equation} 
where: $\ell_q=\ell+P/3$, $k_q=k+P/3$, $\ell_{qq}=-\ell+ 2P/3$, 
$k_{qq}=-k+2P/3$ and the superscript ``T'' denotes matrix transpose; and
\begin{eqnarray}
\nonumber
\lefteqn{({\cal M}_{01})^j_\nu= {\tt t}^j \,
\Gamma_\mu^{1^+}\!(k_q-\ell_{qq}/2;\ell_{qq})} \\
&& \times 
S^{\rm T}(\ell_{qq}-k_q)\,\bar\Gamma^{0^+}\!(\ell_q-k_{qq}/2;-k_{qq})\, 
S(\ell_q)\,\Delta^{1^+}_{\mu\nu}(\ell_{qq}) \,, \label{calM01} \\ 
\nonumber \lefteqn{({\cal M}_{10})^i_\mu = 
\Gamma^{0^+}\!(k_q-\ell_{qq}/2;\ell_{qq})\, 
}\\ 
&&\times S^{\rm T}(\ell_{qq}-k_q)\,{\tt t}^i\, \bar\Gamma_\mu^{1^+}\!(\ell_q-k_{qq}/2;-k_{qq})\, 
S(\ell_q)\,\Delta^{0^+}(\ell_{qq}) \,,\\ 
\nonumber \lefteqn{({\cal M}_{11})^{ij}_{\mu\nu} = {\tt t}^j\, 
\Gamma_\rho^{1^+}\!(k_q-\ell_{qq}/2;\ell_{qq})}\\ 
&&\times \, S^{\rm T}(\ell_{qq}-k_q)\,{\tt t}^i\, \bar\Gamma^{1^+}_\mu\!(\ell_q-k_{qq}/2;-k_{qq})\, 
S(\ell_q)\,\Delta^{1^+}_{\rho\nu}(\ell_{qq}) \,. \label{calM11} 
\end{eqnarray} 

\subsection{Kernel of the Faddeev equation}
\label{completing}
To complete the Faddeev equations, Eq.\,(\ref{FEone}), one must specify the dressed-quark propagator, the diquark Bethe-Salpeter amplitudes and the diquark propagators.

\subsubsection{Dressed-quark propagator} 
\label{subsubsec:S} 
The dressed-quark propagator has the general form 
\begin{eqnarray} 
S(p) & = & -i \gamma\cdot p\, \sigma_V(p^2) + \sigma_S(p^2) = 1/[i\gamma\cdot p\, A(p^2) + B(p^2)]\label{SpAB} 
\end{eqnarray}
and can be obtained from QCD's gap equation; namely, the Dyson-Schwinger equation for the dressed-quark self-energy \cite{Roberts:1994dr}.  The gap equation has been much studied and features of its solution elucidated.  Indeed, it is a longstanding prediction of DSE studies in QCD that for light-quarks the wave function renormalisation and dressed-quark mass: 
\begin{equation} 
\label{ZMdef}
Z(p^2)=1/A(p^2)\,,\;M(p^2)=B(p^2)/A(p^2)\,, 
\end{equation} 
respectively, receive strong momentum-dependent corrections at infrared momenta \cite{Roberts:1994dr,Lane:1974he,Politzer:1976tv}: $Z(p^2)$ is suppressed and $M(p^2)$ enhanced.  These features are an expression of dynamical chiral symmetry breaking (DCSB) and, plausibly, of confinement \cite{Roberts:2007ji}.  The enhancement of $M(p^2)$ is central to the appearance of a constituent-quark mass-scale and an existential prerequisite for Goldstone modes.  The mass function evolves with increasing $p^2$ to reproduce the asymptotic behaviour familiar from perturbative analyses, and that behaviour is unambiguously evident for $p^2 \gtrsim 10\,$GeV$^2$ \cite{Ivanov:1998ms}.  These DSE predictions are confirmed in numerical simulations of lattice-regularised QCD \cite{bowman}, and the conditions have been explored under which pointwise agreement between DSE results and lattice simulations may be obtained \cite{bhagwat,bhagwat2,bhagwattandy}.

The impact of this infrared dressing on hadron phenomena has long been emphasised \cite{cdrpion} and, while numerical solutions of the quark DSE are now readily obtained, the utility of an algebraic form for $S(p)$ when calculations require the evaluation of numerous multidimensional integrals is self-evident.  An efficacious parametrisation 
of $S(p)$, which exhibits the features described above, has been used 
extensively in hadron studies \cite{Roberts:2007jh}.  It is expressed via
\begin{eqnarray} 
\bar\sigma_S(x) & =&  2\,\bar m \,{\cal F}(2 (x+\bar m^2)) + {\cal
F}(b_1 x) \,{\cal F}(b_3 x) \,  
\left[b_0 + b_2 {\cal F}(\epsilon x)\right]\,,\label{ssm} \\ 
\label{svm} \bar\sigma_V(x) & = & \frac{1}{x+\bar m^2}\, \left[ 1 - {\cal F}(2 (x+\bar m^2))\right]\,, 
\end{eqnarray}
with $x=p^2/\lambda^2$, $\bar m$ = $m/\lambda$, 
\begin{equation}
\label{defcalF}
{\cal F}(x)= \frac{1-\mbox{\rm e}^{-x}}{x}  \,, 
\end{equation}
$\bar\sigma_S(x) = \lambda\,\sigma_S(p^2)$ and $\bar\sigma_V(x) =
\lambda^2\,\sigma_V(p^2)$.  The mass-scale, $\lambda=0.566\,$GeV, and
parameter values\footnote{The parameters $b_{0,1,2,3}$ are assumed to be $m$-independent.  In the current application, that is possibly a weakness of the parametrisation.  For example, it leads to a constituent-quark $\sigma$-term that is $30$\% smaller than that obtained from the solution of a well-constrained gap equation \protect\cite{Flambaum:2005kc}.  $\epsilon=10^{-4}$ in Eq.\ (\ref{ssm}) acts only to
decouple the large- and intermediate-$p^2$ domains.}
\begin{equation} 
\label{tableA} 
\begin{array}{ccccc} 
   \bar m& b_0 & b_1 & b_2 & b_3 \\\hline 
   0.00897 & 0.131 & 2.90 & 0.603 & 0.185 
\end{array}\;, 
\end{equation} 
were fixed in a least-squares fit to light-meson observables \cite{mark,valencedistn}.  The dimensionless $u=d$ current-quark mass in Eq.~(\ref{tableA}) corresponds to
\begin{equation} 
\label{mcq}
m=5.08\,{\rm MeV} = :m^{\rm phys}\,. 
\end{equation} 
The parametrisation yields a Euclidean constituent-quark mass
\begin{equation} 
\label{MEq} M_{u,d}^E = 0.33\,{\rm GeV}, 
\end{equation} 
defined as the solution of $p^2=M^2(p^2)$.  

The ratio $M^E/m = 65$ is one expression of DCSB in the parametrisation of $S(p)$.  It emphasises the dramatic enhancement of the dressed-quark mass function at infrared momenta. Another is the chiral-limit vacuum quark condensate \cite{cdrpion}
\begin{equation}
\label{qbqparam}
-\langle \bar q q \rangle_\zeta^0 = \lambda^3 \, \frac{3}{4\pi^2}\, \frac{b_0}{b_1 b_3} \, \ln \frac{\zeta^2}{\Lambda_{\rm QCD}^2}
\end{equation}
which assumes the value ($\Lambda_{\rm QCD} = 0.2\,$GeV)
\begin{equation}
-\langle \bar q q \rangle_{\zeta=1\,{\rm GeV}}^0 = (0.221 \,{\rm GeV})^3.
\end{equation}
A detailed discussion of the vacuum quark condensate in QCD can be found in Ref.\,\cite{Langfeld:2003ye,Chang:2006bm}

An exact formula for pseudoscalar meson masses was derived in Ref.\,\cite{Maris:1997hd} and in the present context it can be expressed for the pion as
\begin{equation}
\label{inpion}
f_\pi^2 m_\pi^2 = - 2 m \langle \bar q q \rangle_{1\,{\rm GeV}}^\pi\,,
\end{equation}
where $\langle \bar q q \rangle_{1\,{\rm GeV}}^\pi$ is an in-pion condensate \cite{Maris:1997tm}.  In order to calculate this quantity and $f_\pi$ one needs the pion's Bethe-Salpeter amplitude.  An \textit{Ansatz} compatible with the parametrisation of the dressed-quark propagator described above is explained in Ref.\,\cite{valencedistn}, and together they yield 
\begin{equation}
- \langle \bar q q \rangle_{1\,{\rm GeV}}^\pi = (0.250\,{\rm GeV})^3\,,\; 
f_\pi = 0.090 \,{\rm GeV}\,,\;
m_\pi = 0.140\,{\rm GeV}\,.
\end{equation}

\subsubsection{Diquark Bethe-Salpeter amplitudes}
\label{qqBSA}
The rainbow-ladder DSE truncation yields asymptotic diquark states in the strong interaction spectrum.  Such states are not observed and their appearance is an artefact of the truncation.  Higher-order terms in the quark-quark scattering kernel, whose analogue in the quark-antiquark channel do not much affect the properties of vector and flavour non-singlet pseudoscalar mesons, ensure that QCD's quark-quark scattering matrix does not exhibit singularities which correspond to asymptotic diquark states~\cite{mandarvertex}.  Nevertheless, studies with kernels that do not produce diquark bound states, do support a physical interpretation of the masses, $m_{(qq)_{J^P}}$, obtained using the rainbow-ladder truncation: the quantity $l_{(qq)_{J^P}}=1/m_{(qq)_{J^P}}$ may be interpreted as a range over which the diquark correlation can persist inside a baryon.  These observations motivate an {\it Ansatz} for the quark-quark scattering matrix that is employed in deriving the Faddeev equation: 
\begin{equation} 
[M_{qq}(k,q;K)]_{rs}^{tu} = \sum_{J^P=0^+,1^+,\ldots} \bar\Gamma^{J^P}\!(k;-K)\, \Delta^{J^P}\!(K) \, \Gamma^{J^P}\!(q;K)\,. \label{AnsatzMqq} 
\end{equation}  

One means of specifying the $\Gamma^{J^P}\!\!$ in Eq.\,(\ref{AnsatzMqq}) is to employ the solutions of a rainbow-ladder quark-quark Bethe-Salpeter equation (BSE), as e.g.\ in Refs.\ \cite{Maris:2002yu,Maris:2004bp}.  Using the properties of the Gell-Mann matrices one finds easily that $\Gamma^{J^P}_C:= \Gamma^{J^P}C^\dagger$ satisfies exactly the same equation as the $J^{-P}$ colour-singlet meson {\it but} for a halving of the coupling strength \cite{Cahill:1987qr}.  This makes clear that the interaction in the ${\bar 3_c}$ $(qq)$ channel is strong and attractive.  

A solution of the BSE equation requires a simultaneous solution of the quark-DSE.  However, since we have already chosen to simplify the calculations by parametrising $S(p)$, we also employ that expedient with $\Gamma^{J^P}\!$, using the following one-parameter forms: 
\begin{eqnarray} 
\label{Gamma0p} \Gamma^{0^+}(k;K) &=& \frac{1}{{\cal N}^{0^+}} \, 
H^a\,C i\gamma_5\, i\tau_2\, {\cal F}(k^2/\omega_{0^+}^2) \,, \\ 
\label{Gamma1p} {\tt t}^i \Gamma^{1^+}_\mu (k;K) &=& \frac{1}{{\cal N}^{1^+}}\, 
H^a\,i\gamma_\mu C\,{\tt t}^i\, {\cal F}(k^2/\omega_{1^+}^2)\,, 
\end{eqnarray} 
with the normalisation, ${\cal N}^{J^P}\!$, fixed by requiring 
\begin{eqnarray}
\label{BSEnorm} 
2 \,K_\mu & = & 
\left[ \frac{\partial}{\partial Q_\mu} \Pi(K,Q) \right]_{Q=K}^{{K^2=-m_{J^P}^2}},\\
\Pi(K,Q) & = & tr\!\! \int\!\! 
\frac{d^4 q}{(2\pi)^4}\, \bar\Gamma(q;-K) \, S(q+Q/2) \, \Gamma(q;K) \, S^{\rm T}(-q+Q/2) .
\end{eqnarray}

The {\it Ans\"atze} of Eqs.\,(\ref{Gamma0p}), (\ref{Gamma1p}) retain only that single Dirac-amplitude which would represent a point particle with the given quantum numbers in a local Lagrangian density.  They are usually the dominant amplitudes in a solution of the rainbow-ladder BSE for the lowest mass $J^P$ diquarks \cite{Burden:1996nh,Maris:2002yu} and mesons \cite{Maris:1997tm,Maris:1999nt,Maris:1999bh}.

\subsubsection{Diquark propagators}
\label{qqprop}
Solving for the quark-quark scattering matrix using the rainbow-ladder truncation yields free particle propagators for $\Delta^{J^P}$ in 
Eq.~(\ref{AnsatzMqq}).  As already noted, however, higher-order contributions 
remedy that defect, eliminating asymptotic diquark states from the spectrum.  The attendant modification of $\Delta^{J^P}$ can be modelled efficiently by simple functions that are free-particle-like at spacelike momenta but pole-free on the timelike axis \cite{mandarvertex}; namely,\footnote{These forms satisfy a sufficient condition for confinement because of the associated violation of reflection positivity.  See Sect.~2 of Ref.\,\cite{Roberts:2007ji} for a brief discussion.}
\begin{eqnarray} 
\Delta^{0^+}(K) & = & \frac{1}{m_{0^+}^2}\,{\cal F}(K^2/\omega_{0^+}^2)\,,\\ 
\Delta^{1^+}_{\mu\nu}(K) & = & 
\left(\delta_{\mu\nu} + \frac{K_\mu K_\nu}{m_{1^+}^2}\right) \, \frac{1}{m_{1^+}^2}\, {\cal F}(K^2/\omega_{1^+}^2) \,,
\end{eqnarray} 
where the two parameters $m_{J^P}$ are diquark pseudoparticle masses and 
$\omega_{J^P}$ are widths characterising $\Gamma^{J^P}\!$.  Herein we require additionally that
\begin{equation}
\label{DQPropConstr}
\left. \frac{d}{d K^2}\,\left(\frac{1}{m_{J^P}^2}\,{\cal F}(K^2/\omega_{J^P}^2)\right)^{-1} \right|_{K^2=0}\! = 1 \; \Rightarrow \; \omega_{J^P}^2 = \sfrac{1}{2}\,m_{J^P}^2\,,
\end{equation} 
which is a normalisation that accentuates the free-particle-like propagation characteristics of the diquarks {\it within} the hadron. 

\section{Euclidean Conventions}
\label{App:EM} 
In our Euclidean formulation: 
\begin{equation} 
p\cdot q=\sum_{i=1}^4 p_i q_i\,; 
\end{equation} 
\begin{equation}
\{\gamma_\mu,\gamma_\nu\}=2\,\delta_{\mu\nu}\,;\; 
\gamma_\mu^\dagger = \gamma_\mu\,;\; 
\sigma_{\mu\nu}= \sfrac{i}{2}[\gamma_\mu,\gamma_\nu]\,; \;
{\rm tr}_[\gamma_5\gamma_\mu\gamma_\nu\gamma_\rho\gamma_\sigma]= 
-4\,\epsilon_{\mu\nu\rho\sigma}\,, \epsilon_{1234}= 1\,.  
\end{equation}

A positive energy spinor satisfies 
\begin{equation} 
\bar u(P,s)\, (i \gamma\cdot P + M) = 0 = (i\gamma\cdot P + M)\, u(P,s)\,, 
\end{equation} 
where $s=\pm$ is the spin label.  It is normalised: 
\begin{equation} 
\bar u(P,s) \, u(P,s) = 2 M 
\end{equation} 
and may be expressed explicitly: 
\begin{equation} 
u(P,s) = \sqrt{M- i {\cal E}}\left( 
\begin{array}{l} 
\chi_s\\ 
\displaystyle \frac{\vec{\sigma}\cdot \vec{P}}{M - i {\cal E}} \chi_s 
\end{array} 
\right)\,, 
\end{equation} 
with ${\cal E} = i \sqrt{\vec{P}^2 + M^2}$, 
\begin{equation} 
\chi_+ = \left( \begin{array}{c} 1 \\ 0  \end{array}\right)\,,\; 
\chi_- = \left( \begin{array}{c} 0\\ 1  \end{array}\right)\,. 
\end{equation} 
For the free-particle spinor, $\bar u(P,s)= u(P,s)^\dagger \gamma_4$. 
 
The spinor can be used to construct a positive energy projection operator: 
\begin{equation} 
\label{Lplus} \Lambda_+(P):= \frac{1}{2 M}\,\sum_{s=\pm} \, u(P,s) \, \bar 
u(P,s) = \frac{1}{2M} \left( -i \gamma\cdot P + M\right). 
\end{equation} 
 
A negative energy spinor satisfies 
\begin{equation} 
\bar v(P,s)\,(i\gamma\cdot P - M) = 0 = (i\gamma\cdot P - M) \, v(P,s)\,, 
\end{equation} 
and possesses properties and satisfies constraints obtained via obvious analogy 
with $u(P,s)$. 
 
A charge-conjugated Bethe-Salpeter amplitude is obtained via 
\begin{equation} 
\label{chargec}
\bar\Gamma(k;P) = C^\dagger \, \Gamma(-k;P)^{\rm T}\,C\,, 
\end{equation} 
where ``T'' denotes a transposing of all matrix indices and 
$C=\gamma_2\gamma_4$ is the charge conjugation matrix, $C^\dagger=-C$. 

\section{Nucleon-Photon Vertex}
\label{NPVertex}
In order to explicate the vertex depicted in Fig.\,\ref{vertex} we write the scalar and axial-vector components of the nucleons' Faddeev amplitudes in the form [cf.\ Eq.\,(\ref{FEone})]
\begin{equation}
\label{NucWF}
\Psi(k;P) = \left[
\begin{array}{l}
\Psi^0(k;P)\\
\Psi_{\mu}^{i}(k;P)
\end{array}
\right]
= \left[ 
\begin{array}{l}
\mathcal{S}(k;P) u(P)\\
\mathcal{A}_{\mu}^{i}(k;P)u(P)
\end{array}
\right],
\qquad i=1,\ldots,4\,.
\end{equation}
For explicit calculations, we work in the Breit frame: $P_\mu=P_\mu^{BF}-Q_\mu /2$, $P'_\mu=P_\mu^{BF}+Q_\mu /2$ and $P_\mu^{BF}=(0,0,0,i\sqrt{M_n^2+Q^2/4})$, and write the electromagnetic current matrix element as [cf.\ Eq.\,(\ref{Jnucleon})]
\begin{eqnarray}
\label{ABcurrent}
\left\langle P' | \hat{J}_{\mu}^{em} | P \right\rangle
&=& \Lambda^+(P')
\left[ \gamma_\mu G_E + M_n \frac{P_{\mu}^{BF}}{P_{BF}^{2}}
(G_E-G_M) \right] \Lambda^+(P)
\\
&=& \int \frac{d^4 p}{(2\pi)^4}\,\frac{d^4 k}{(2\pi)^4}\,
\bar{\Psi}(-p,P') J_{\mu}^{em}(p,P';k,P) \Psi(k,P)\,.
\end{eqnarray}
In Fig.\,\ref{vertex} we have broken the current, $J_{\mu}^{em}(p,P';k,P)$,  into a sum of six terms, each of which we subsequently make precise.  NB.\ Diagrams~1, 2 and 4 are one-loop integrals, which we evaluate by Gau{\ss}ian quadrature.  The remainder, Diagrams 3, 5 and 6, are two-loop integrals, for whose evaluation Monte-Carlo methods are employed.  

\subsection{Diagram~1}
\label{Diag1}
This represents the photon coupling directly to the bystander quark. It is expressed as \begin{eqnarray}
\label{B1}
J_{\mu}^{qu} &=&  S(p_q) \hat{\Gamma}_{\mu}^{qu}(p_q;k_q) S(k_q) 
\left(\Delta^{0^+}(k_s) + \Delta^{1^+}(k_s) \right)
(2\pi)^4 \delta^4(p-k-\hat{\eta}Q)\,,
\end{eqnarray}
where $\hat\Gamma_{\mu}^{qu}(p_q;k_q)= Q_q \, \Gamma_{\mu}(p_q;k_q)$, with $Q_q={\rm diag}[2/3,-1/3]$ being the quark electric charge matrix, and $\Gamma_{\mu}(p_q;k_q)$ is the dressed-quark-photon vertex.  In Eq.\,(\ref{B1}) the momenta are
\begin{eqnarray}
\label{etavalue}
\begin{array}{lc@{\qquad}l}
k_q=\eta P+k\,, & & p_q=\eta P'+p\,, \\
k_d=\hat{\eta}P-k\,, & & p_d=\hat{\eta}P'-p\,,
\end{array}
\end{eqnarray}
with $\eta + \hat{\eta}=1$.  The results reported herein were obtained with $\eta=1/3$, which provides a single quark with one-third of the baryon's total momentum and is thus a natural choice.  Notably, as our approach is manifestly Poincar\'e covariant, the precise value is immaterial so long as the numerical methods preserve that covariance.  Calculations converge most quickly with the natural choice. 

It is a necessary condition for current conservation that the quark-photon vertex satisfy the Ward-Takahashi identity:
\begin{equation}
\label{vwti}
Q_\mu \, i\Gamma_\mu(\ell_1,\ell_2) = S^{-1}(\ell_1) - S^{-1}(\ell_2)\,,
\end{equation}
where $Q=\ell_1-\ell_2$ is the photon momentum flowing into the vertex.  Since the quark is dressed, Sec.\,\ref{subsubsec:S}, the vertex is not bare; i.e., $\Gamma_\mu(\ell_1,\ell_2) \neq \gamma_\mu$.  It can be obtained by solving an inhomogeneous Bethe-Salpeter equation, which was the procedure adopted in the DSE calculation that successfully predicted the electromagnetic pion form factor \cite{Maris:2000sk,Maris:1999bh}.  However, since we have parametrised $S(p)$, we follow Ref.~\cite{cdrpion} and write \cite{bc80}
\begin{equation}
\label{bcvtx}
i\Gamma_\mu(\ell_1,\ell_2)  =  
i\Sigma_A(\ell_1^2,\ell_2^2)\,\gamma_\mu +
2 k_\mu \left[i\gamma\cdot k_\mu \,
\Delta_A(\ell_1^2,\ell_2^2) + \Delta_B(\ell_1^2,\ell_2^2)\right] \!;
\end{equation}
with $k= (\ell_1+\ell_2)/2$, $Q=(\ell_1-\ell_2)$ and
\begin{equation}
\Sigma_F(\ell_1^2,\ell_2^2) = \sfrac{1}{2}\,[F(\ell_1^2)+F(\ell_2^2)]\,,\;
\Delta_F(\ell_1^2,\ell_2^2) =
\frac{F(\ell_1^2)-F(\ell_2^2)}{\ell_1^2-\ell_2^2}\,,
\label{DeltaF}
\end{equation}
where $F= A, B$; viz., the scalar functions in Eq.\,(\ref{SpAB}).  It is
critical that $\Gamma_\mu$ in Eq.\ (\ref{bcvtx}) satisfies Eq.\ (\ref{vwti})
and very useful that it is completely determined by the dressed-quark
propagator.  

\subsection{Diagram~2} 
\label{dqff}
This figure depicts the photon coupling directly to a diquark correlation.  It is expressed as 
\begin{eqnarray}
\label{B2}
J_{\mu}^{dq} &=& \Delta^i(p_{d}) 
\left[ \hat{\Gamma}_{\mu}^{dq}(p_{d};k_{d}) \right]^{i j} 
\Delta^{j}(k_{d}) S(k_q)
(2\pi)^4 \delta^4(p-k+\eta Q)\,
\end{eqnarray}
with $ [\hat{\Gamma}_{\mu}^{dq}(p_{d};k_{d})]^{i j}={\rm diag}[Q_{0^+} \Gamma_\mu^{0^+},Q_{1^+}\Gamma_\mu^{1^+}] $, where $Q_{0^+}=1/3$ and $\Gamma_\mu^{0^+}$ is given in Eq.\,(\ref{Gamma0plus}), and $Q_{1^+}={\rm diag}[4/3,1/3,-2/3]$ with $\Gamma_\mu^{1^+}$ given in Eq.\,(\ref{AXDQGam}).  Naturally, the diquark propagators match the line to which they are attached. 

In the case of a scalar correlation, the general form of the diquark-photon vertex is
\begin{equation}
\Gamma_\mu^{0^+}(\ell_1,\ell_2) = 2\, k_\mu\, f_+(k^2,k\cdot Q,Q^2) + Q_\mu  \, f_-(k^2,k\cdot Q,Q^2)\,,
\end{equation}
and it must satisfy the Ward-Takahashi identity: 
\begin{equation}
\label{VWTI0}
Q_\mu \,\Gamma_\mu^{0^+}(\ell_1,\ell_2) = \Pi^{0^+}(\ell_1^2)  - \Pi^{0^+}(\ell_2^2)\,,\; \Pi^{J^P}(\ell^2) = \{\Delta^{J^P}(\ell^2)\}^{-1}.
\end{equation} 
The evaluation of scalar diquark elastic electromagnetic form factors in Ref.\,\cite{Maris:2004bp} is a first step toward calculating this vertex.  However, in providing only an on-shell component, it is insufficient for our requirements.  We therefore adapt Eq.\,(\ref{bcvtx}) to this case and write
\begin{equation}
\label{Gamma0plus}
\Gamma_\mu^{0^+}(\ell_1,\ell_2) =  k_\mu\,
\Delta_{\Pi^{0^+}}(\ell_1^2,\ell_2^2)\,,
%
\end{equation}  
with the definition of $\Delta_{\Pi^{0^+}}(\ell_1^2,\ell_2^2)$ apparent from Eq.\,(\ref{DeltaF}).  Equation~(\ref{Gamma0plus}) is the minimal \textit{Ansatz} that: satisfies Eq.\,(\ref{VWTI0}); is completely determined by quantities introduced already; and is free of kinematic singularities.  It implements $f_- \equiv 0$, which is a requirement for elastic form factors, and guarantees a valid normalisation of electric charge; viz., 
\begin{equation}
\lim_{\ell^\prime\to \ell} \Gamma_\mu^{0^+}(\ell^\prime,\ell) = 2 \, \ell_{\mu} \, \frac{d}{d\ell^2}\, \Pi^{0^+}(\ell^2) \stackrel{\ell^2\sim 0}{=} 2 \, \ell_{\mu}\,,
\end{equation}
owing to Eq.\,(\ref{DQPropConstr}).  NB.\ We have factored the fractional diquark charge, which therefore appears subsequently in our calculations as a simple multiplicative factor. 

For the case in which the struck diquark correlation is axial-vector and the scattering is elastic, the vertex assumes the form \cite{HawesPichowsky99}:\,\footnote{If the scattering is inelastic the general form of the vertex involves eight scalar functions \protect\cite{Salam64}.  For simplicity, we ignore the additional structure in this \textit{Ansatz}.  
}
\begin{equation}
\label{AXDQGam}
\Gamma^{1^+}_{\mu\alpha\beta}(\ell_1,\ell_2) 
= -\sum_{i=1}^{3} \Gamma^{\rm [i]}_{\mu\alpha\beta}(\ell_1,\ell_2)\,,
\end{equation}
with ($T_{\alpha\beta}(\ell) = \delta_{\alpha\beta} - \ell_\alpha \ell_\beta/\ell^2$)
\begin{eqnarray}
\label{AXDQGam1}
\Gamma^{\rm [1]}_{\mu\alpha\beta}(\ell_1,\ell_2) 
&=& (\ell_1+\ell_2)_\mu \, T_{\alpha\lambda}(\ell_1) \, T_{\lambda\beta}(\ell_2)\; F_1(\ell_1^2,\ell_2^2)\,,
\\
\label{AXDQGam2}
\Gamma^{\rm [2]}_{\mu\alpha\beta}(\ell_1,\ell_2)
&=& \left[ T_{\mu\alpha}(\ell_1)\, T_{\beta\rho}(\ell_2) \, \ell_{1 \rho}
+ T_{\mu\beta}(\ell_2) \, T_{\alpha\rho}(\ell_1) \, \ell_{2\rho} \right] F_{2}(\ell_1^2,\ell_2^2) \,,
\\ 
\label{AXDQGam3}
\Gamma^{\rm [3]}_{\mu\alpha\beta}(\ell_1,\ell_2)
&=& -\frac{1}{2 m_{1^+}^2}\, (\ell_1+\ell_2)_\mu\, T_{\alpha\rho}(\ell_1)\, \ell_{2 \rho}
\, T_{\beta\lambda}(\ell_2)\, \ell_{1 \lambda}\; F_{3}(\ell_1^2,\ell_2^2) \,.
\end{eqnarray}
This vertex satisfies:
\begin{equation}
\ell_{1\alpha} \, \Gamma^{1^+}_{\mu\alpha\beta}(\ell_1,\ell_2) = 0 = 
\Gamma^{1^+}_{\mu\alpha\beta}(\ell_1,\ell_2) \, \ell_{2\beta} \,,
\end{equation}
which is a general requirement of the elastic electromagnetic vertex of axial-vector bound states and guarantees that the interaction does not induce a pseudoscalar component in the axial-vector correlation.  We note that the electric, magnetic and quadrupole form factors of an axial-vector bound state are expressed \cite{HawesPichowsky99}
\begin{eqnarray}
\label{GEDQ}
& &
G_{\cal E}^{1^+}(Q^2) = F_1 + \sfrac{2}{3}\, \tau_{1^+}\, 
G_{\cal Q}^{1^+}(Q^2) \,, \; \tau_{1^+} = \frac{Q^2}{ 4 \,m_{1^{+}}^{2}}
\\
\label{GMDQ}
& &
G_{\cal M}^{1^+}(Q^2) = - F_2(Q^2) ~,
\\
& &
\label{GQDQ}
G_{\cal Q}^{1^+}(Q^2) = F_1(Q^2) + F_2(Q^2) + \left( 1 + \tau_{1^+}\right) F_3(Q^2) \,.
\end{eqnarray}

Owing to the fact that $\Gamma^{J^P}_C:= \Gamma^{J^P}C^\dagger$ satisfies exactly the same Bethe-Salpeter equation as the $J^{-P}$ colour-singlet meson {\it but} for a halving of the coupling strength, the vector meson form factor calculation in Ref.\,\cite{Bhagwat:2006pu} might become useful as a guide in understanding the form factors in Eqs.\,(\ref{AXDQGam})--(\ref{AXDQGam3}).  However, in providing only an on-shell component, that information is insufficient for our requirements.  Hence, we employ the following \textit{Ans\"atze}:
\begin{eqnarray}
\label{AnsatzF1}
F_{1}(\ell_1^2,\ell_2^2) &=& \Delta_{\Pi^{1^+}}(\ell_1^2,\ell_2^2)\,, \\
\label{AnsatzF2}
F_{2}(\ell_1^2,\ell_2^2) &=& -\, F_{1} + 
(1-\tau_{1^+}) \,( \tau_{1^+} F_{1}+1 - \mu_{1^+})\, d(\tau_{1^+}) \\
\label{AnsatzF3}
F_{3}(\ell_1^2,\ell_2^2) &=& -\,(\chi_{1^+}\,(1- \tau_{1^+})\,d(\tau_{1^+})+F_1 + F_2)\, d(\tau_{1^+})\,,
\end{eqnarray}
with $d(x)=1/(1+x)^2$.  This construction ensures a valid electric charge normalisation for the axial-vector correlation; viz., 
\begin{equation}
\lim_{\ell^\prime \to\ell} \, \Gamma^{1^+}_{\mu\alpha\beta}(\ell^\prime,\ell) = T_{\alpha\beta}(\ell) \,\frac{d}{d\ell^2}\, \Pi^{1^+}(\ell^2) 
\stackrel{\ell^2\sim 0}{=}  T_{\alpha\beta}(\ell) \,2 \,\ell_{ \mu}\,,
\end{equation}
owing to Eq.\,(\ref{DQPropConstr}), and current conservation 
\begin{equation}
\lim_{\ell_2\to\ell_1} \, Q_\mu \Gamma^{1^+}_{\mu\alpha\beta}(\ell_1,\ell_2) = 0\,.
\end{equation}
The diquark's static electromagnetic properties follow: 
\begin{equation}
\label{pointp}
G_{\cal E}^{1^+}(0) = 1\,,\;
G_{\cal M}^{1^+}(0) = \mu_{1^+}\,,\;
G_{\cal Q}^{1^+}(0) = -\chi_{1^+}\,.
\end{equation}
For an on-shell or pointlike axial-vector: $\mu_{1^+}=2$; and $\chi_{1^+}=1$, which corresponds to an oblate charge distribution.  In addition, Eqs.\,(\ref{AXDQGam})--(\ref{AXDQGam3}) with Eqs.\,(\ref{AnsatzF1})--(\ref{AnsatzF3}) realise the constraints of Ref.\,\cite{brodskyhiller92}; namely, independent of the values of $\mu_{1^+}$ \& $\chi_{1^+}$, the form factors assume the ratios
\begin{equation}
\label{pQCDavdq}
G_{\cal E}^{1^+}(Q^2): G_{\cal M}^{1^+}(Q^2): G_{\cal Q}^{1^+}(Q^2)
\stackrel{Q^2\to \infty}{=} (1 - \sfrac{2}{3} \tau_{1^+}) : 2 : - 1 \,.
\end{equation}

We note that within a nucleon the diquark correlation is not on-shell.  Hence, in contrast with Ref.\,\cite{Alkofer:2004yf}, herein we do not assume that point-particle values for the magnetic and quadrupole moments in Eqs.\,(\ref{pointp}) serve as a good reference point.  For the processes described by Fig.\,\ref{vertex}, the values can be much smaller in magnitude \cite{Mineo:2002bg}, as we find in Table~\ref{magmoms}.

\subsection{Diagram~3}
This image depicts a photon coupling to the quark that is exchanged as one diquark breaks up and another is formed.  It is expressed as
\begin{equation}
\label{B3}
J_{\mu}^{ex} = -\frac{1}{2} S(k_{q}) \Delta^{i}(k_{d})
\Gamma^{i}(p_1,k_{d})
S^T(q) \hat{\Gamma}_{\mu}^{quT}(q',q) S^T(q')
\bar{\Gamma}^{jT}(p'_2,p_{d})
\Delta^j(p_{d}) S(p_{q})\,,
\end{equation}
wherein the vertex $\hat{\Gamma}_{\mu}^{qu}$ appeared in Eq.\,(\ref{B1}).  While this is the first two-loop diagram we have described, no new elements appear in its specification: the dressed-quark-photon vertex was discussed in Sec.~\ref{Diag1}.  In Eq.\,(\ref{B3}) the momenta are 
\begin{eqnarray}
\begin{array}{lc@{\qquad}l}
q = \hat{\eta}P-\eta P'-p-k\,, & & q' = \hat{\eta}P'-\eta P-p-k \,,\\
p_1 = (p_q-q)/2\,,& & p'_2 = (-k_q+q')/2 \,,\\
p'_1 = (p_q-q')/2\,, & & p_2 = (-k_q+q)/2 \,.
\end{array}
\end{eqnarray}

It is noteworthy that the process of quark exchange provides the attraction necessary in the Faddeev equation to bind the nucleon.  It also guarantees that the Faddeev amplitude has the correct antisymmetry under the exchange of any two dressed-quarks.  This key feature is absent in models with elementary (noncomposite) diquarks.  The full contribution is obtained by summing over the superscripts $i,j$, which can each take the values $0^+$, $1^+$.  

\subsection{Diagram~4}
\label{kTsec}
This differs from Diagram~2 in expressing the contribution to the nucleons' form factors owing to an electromagnetically induced transition between scalar and axial-vector diquarks.  The explicit expression is given by Eq.\,(\ref{B2}) with $[\hat{\Gamma}_{\mu}^{dq}(p_{d};k_{d})]^{i= j}=0$, and $[\hat{\Gamma}_{\mu}^{dq}(p_{d};k_{d})]^{1,2}=\Gamma_{SA}$ and $[\hat{\Gamma}_{\mu}^{dq}(p_{d};k_{d})]^{2,1}=\Gamma_{AS}$.  This transition vertex is a rank-2 pseudotensor, kindred to the matrix element describing the $\rho\, \gamma^\ast \pi^0$  transition \cite{maristransition}, and can therefore be expressed 
\begin{equation}
\label{SAPhotVertex}
\Gamma_{SA}^{\gamma\alpha}(\ell_1,\ell_2) = -\Gamma_{AS}^{\gamma\alpha}(\ell_1,\ell_2) 
= \frac{i}{M_N} \, {\cal T}(\ell_1,\ell_2) \, \varepsilon_{\gamma\alpha\rho\lambda}\ell_{1\rho} \ell_{2 \lambda}\,,
\end{equation}
where $\gamma$, $\alpha$ are, respectively, the vector indices of the photon and axial-vector diquark.  For simplicity we proceed under the assumption that
\begin{equation}
\label{calTvalue}
{\cal T}(\ell_1,\ell_2) = \kappa_{\cal T}\,;
\end{equation}
viz., a constant, for which a typical on-shell value is 
$\kappa_{\cal T} \sim 2$ \cite{Oettel:2000jj}.  
However, as with $\mu_{1^+}$ and $\chi_{1^+}$, we recognise herein that this value is not a useful reference point because, for the processes described by Fig.\,\ref{vertex}, $\kappa_{\cal T}$ can be much smaller in magnitude.

In the nucleons' rest frame, a conspicuous piece of the Faddeev amplitude that describes an axial-vector diquark inside the bound state can be characterised as containing a bystander quark whose spin is antiparallel to that of the nucleon, with the axial-vector diquark's spin parallel.  The interaction pictured in this diagram does not affect the bystander quark but the transformation of an axial-vector diquark into a scalar effects a flip of the quark spin within the correlation.  After this transformation, the spin of the nucleon must be formed by summing the spin of the bystander quark, which is still aligned antiparallel to that of the nucleon, and the orbital angular momentum between that quark and the scalar diquark.  This argument, while not sophisticated, does motivate an expectation that Diagram~4 will impact upon the nucleons' magnetic form factors.\footnote{Another component of the amplitude has the bystander quark's spin parallel to that of the nucleon while the axial-vector diquark's is antiparallel: this $q^\uparrow \oplus (qq)_{1^+}^{\downarrow} $ system has one unit of angular momentum.  That momentum is absent in the $q^\uparrow \oplus (qq)_{0^+}$ system.  Other combinations also contribute via Diagram~3 but all mediated processes inevitably require a modification of spin and/or angular momentum.  An analysis of the contribution from quark orbital angular momentum to a nucleon's spin is presented in Ref.\,\protect\cite{Cloet:2007pi}.}

\subsection{Diagrams~5 \& 6}
\label{X56}
These two-loop diagrams are the so-called ``seagull'' terms, which appear as partners to Diagram~3 and arise because binding in the nucleons' Faddeev equations is effected by the exchange of \textit{nonpointlike} diquark correlations \cite{Oettel:1999gc}.  The explicit expression for their contribution to the nucleons' form factors is 
\begin{eqnarray}
\label{B5}
J_{\mu}^{sg} &=& \frac{1}{2} S(k_{q}) \Delta^{i}(k_{d}) 
\left( X_{\mu}^{i}(p_q,q',k_d) S^T(q')
\bar{\Gamma}^{jT}(p'_2,p_{d})
\right.
\nonumber\\
& & -
\left. 
\Gamma^{i}(p_1,k_{d}) S^T(q) 
\bar{X}_{\mu}^{j}(-k_q,-q,p_d)
\right) \Delta^{j}(p_{d}) S(p_{q})\,,
\end{eqnarray}
where, again, the superscripts are summed. 
  
The new elements in these diagrams are the couplings of a photon to two dressed-quarks as they either separate from (Diagram~5) or combine to form (Diagram~6) a diquark correlation.  As such they are components of the five point Schwinger function which describes the coupling of a photon to the quark-quark scattering kernel.  This Schwinger function could be calculated, as is evident from the computation of analogous Schwinger functions relevant to meson observables \cite{marisgppp}.  However, such a calculation provides valid input only when a uniform truncation of the DSEs has been employed to calculate each of the elements described hitherto.  We must instead employ an algebraic parametrisation \cite{Oettel:1999gc}, which for Diagram~5 reads
\begin{eqnarray}
\nonumber
X^{J^P}_\mu(k,Q) & =&  e_{\rm by}\,\frac{4 k_\mu- Q_\mu}{4 k\cdot Q - Q^2}\,\left[\Gamma^{J^P}\!(k-Q/2)-\Gamma^{J^P}\!(k)\right]\\
& +& e_{\rm ex}\,\frac{4 k_\mu+ Q_\mu}{4 k\cdot Q + Q^2}\,\left[\Gamma^{J^P}\!(k+Q/2)-\Gamma^{J^P}\!(k)\right], \label{X5}
\end{eqnarray}
with $k$ the relative momentum between the quarks in the initial diquark, $e_{\rm by}$ the electric charge of the quark which becomes the bystander, and $e_{\rm ex}$ the charge of the quark that is reabsorbed into the final diquark.  Diagram~6 has
\begin{eqnarray}
\nonumber
\bar X^{J^P}_\mu(k,Q) & =&  e_{\rm by}\,\frac{4 k_\mu+ Q_\mu}{4 k\cdot Q + Q^2}\,\left[\bar\Gamma^{J^P}\!(k+Q/2)-\bar\Gamma^{J^P}\!(k)\right]\\
& +& e_{\rm ex}\,\frac{4 k_\mu-Q_\mu}{4 k\cdot Q - Q^2}\,\left[\bar\Gamma^{J^P}\!(k-Q/2)-\bar\Gamma^{J^P}\!(k)\right], \label{X6}
\end{eqnarray}
where $\bar\Gamma^{J^P}\!(\ell)$ is the charge-conjugated amplitude, Eq.\,(\ref{chargec}).  Plainly, these terms vanish if the diquark correlation is represented by a momentum-independent Bethe-Salpeter-like amplitude; i.e., the diquark is pointlike.

It is naturally possible to use more complicated \textit{Ans\"atze}.  However, like Eq.\,(\ref{Gamma0plus}), Eqs.\,(\ref{X5}) \& (\ref{X6}) are simple forms, free of kinematic singularities and sufficient to ensure the nucleon-photon vertex satisfies the Ward-Takahashi identity when the composite nucleon is obtained from the Faddeev equation.

\section{Charge Symmetry}
\label{chargesymmetry}
Our analysis assumes $m_u=m_d$.  Hence the only difference between the $u$- and $d$-quarks is their electric charge.  Our equations and computer codes therefore exhibit the following charge symmetry relations:
\begin{equation}
\mu_n^u = -2 \mu_p^d \,,\; \mu_n^d = -\frac{1}{2} \mu_p^u\,,
\end{equation}
where $\mu_n^u$ means the contribution from the $u$-quark to the magnetic moment of the neutron, etc.; and furthermore
\begin{equation}
\delta \mu_n^p = -2 \delta \mu_p^d\,,\; \delta \mu_n^d = -\frac{1}{2} \delta \mu_p^u\,,
\end{equation}
where, in this section, $\delta \mu_n^p$ means the variation in $\mu_n^p$ owing to a small change in current-quark mass.  Using these equations, one obtains
\begin{eqnarray}
\frac{\delta \mu_p}{\mu_p} =
\frac{\delta \mu_p^u + \delta \mu_p^d}{\mu_p^u + \mu_p^d}\,, & &
\frac{\delta \mu_n}{\mu_n}  \frac{\delta \mu_p^u + 4 \delta \mu_p^d}{\mu_p^u + 4 \mu_p^d}\,.
\end{eqnarray}

Our Faddeev equation model yields
\begin{equation}
\label{muud}
\begin{array}{ll}
\mu_p^u= 2.40\,, & \delta \mu_p^u = -0.032\,,~~\\
\mu_p^d = 0.15 \,,& \delta \mu_p^d = ~0.0034\,,
\end{array}
\end{equation}
from which we obtain the results in Tables~\ref{physicalresults} and \ref{dmudm}.

It is interesting to provide a context for the results in Eq.\,(\ref{muud}).  Suppose one were required to reproduce $\mu_p^{q(qq)}$ in Eq.\,(\ref{qcore}) with nonrelativistic pointlike constituent-quarks.  Such quarks have the magnetic moments:
\begin{equation}
\mu_U = 2 \bar \mu_Q \,, \; \mu_D = -1 \bar \mu_Q\,,
\end{equation}
in terms of which 
\begin{equation}
\mu_p^{Q(QQ)} = \frac{4}{3} \mu_U - \frac{1}{3} \mu_D =: \mu_p^U + \mu_p^D\,.
\end{equation}
With $\bar \mu_Q = 0.85$ one reproduces Eq.\,(\ref{qcore}) and finds
\begin{equation}
\label{muUD}
\mu_p^U = 2.27\,, \; \mu_p^D = 0.28\,.
\end{equation}
A comparison between Eqs.\,(\ref{muud}) and (\ref{muUD}) indicates the presence of correlations in our Faddeev amplitude for the nucleon.  Relative to a generic constituent-quark model, they increase the probability for a $u$-quark to have its spin aligned with that of the proton, and markedly decrease that probability for the $d$-quark.


\end{document}